\documentclass[10pt,a4paper]{article}
\usepackage{physmtemplate-article}

\begin{document}


\title{Precipitation-dissolution cycling drives intermittent salt creeping}
\author[1]{Javier Rodr\'{\i}guez-Rodr\'{\i}guez$^{\dag, }$}
\author[2]{Manikuntala Mukhopadhyay$^{\dag, }$}
\author[3,4]{Lijun Thayyil Raju}
\author[4,5]{Detlef Lohse}
\author[2]{Jasper van der Gucht}
\author[2]{Uddalok Sen$^{\ast, }$}
\affil[1]{Departmento de Ingenier\'{\i}a T\'{e}rmica y de Fluidos, Gregorio Mill\'{a}n Institute for Fluid Dynamics, Nanoscience and Industrial Mathematics, Universidad Carlos III de Madrid, 28911 Legan\'{e}s, Spain}
\affil[2]{Physical Chemistry and Soft Matter Group, Wageningen University and Research, 6708 WE Wageningen, The Netherlands}
\affil[3]{Physics of Fluids Department, Faculty of Science and Technology, University of Twente, 7500 AE Enschede, The Netherlands}
\affil[4]{Canon Production Printing BV, 5900 MA Venlo, The Netherlands}
\affil[5]{Max Planck Institute for Dynamics and Self-Organization, 37077 G\"{o}ttingen, Germany}
\def\thefootnote{$\ast$}\footnotetext{{uddalok.sen@wur.nl, ORCID: \href{https://orcid.org/0000-0001-6355-7605}{0000-0001-6355-7605}}}
\def\thefootnote{\dag}\footnotetext{These authors contributed equally to this work.} \def\thefootnote{\arabic{footnote}}
\date{}
\maketitle


\begin{abstract}
	Salt creeping --- the precipitation of salt crystals away from an evaporating liquid interface along surrounding surfaces --- occurs across settings from geology and cultural-heritage weathering to inkjet printing and carbon sequestration. Yet why its dynamics are sometimes smooth and sometimes violently intermittent has remained unexplained. Here we investigate the confined evaporation of salt solutions from a capillary with unidirectional water loss, and show that intermittent salt creeping is driven by a cycle of precipitation and dissolution sustained entirely by liquid transport. Time-resolved measurements reveal that each burst of crystal growth coincides with episodic imbibition of solution into the growing salt deposit on the outer capillary surface, which couples evaporation to crystallization. Combining experiments with a minimal theoretical model, we show that the difference between the supersaturation at which crystals nucleate and the saturation concentration at which they redissolve --- the metastable zone of the salt --- is sufficient to generate this cycle. The same mechanism accounts for the cases in which creeping is not intermittent: where the concentration at the capillary exit never reaches the nucleation threshold the liquid meniscus recedes smoothly, and where it cannot subsequently fall back below saturation the deposit grows continuously, exhibiting self-amplified creeping.  Our results recast salt creeping as a relaxation oscillator, and identify precipitation-dissolution cycling as a generic route to intermittency in evaporating multicomponent fluids. 
\end{abstract}





\section{Introduction}

Evaporation of multicomponent liquids into a gaseous phase is ubiquitous in both natural and technological settings \cite{erbil-2012-advcolloidinterfacesci, lohse-2020-natrevphys, morris-2021-elife, gelderblom-2022-sm, lohse-2022-arfm, hooiveld-2023-jcis, hooiveld-2025-jcis, haessig-2026-smallmeth}. Of particular importance are aqueous salt solutions, which arise in context ranging from biological systems \cite{martinezpuig-2026-jrsi, ganar-2025-natchem} and porous geological media \cite{cooke-1968-nature, rodrigueznavarro-1999-earthsurfprocesslandforms} to industrial processes such as inkjet printing \cite{lohse-2020-natrevphys, lohse-2022-arfm}, desalination \cite{xu-2021-advfunctmater, yang-2024-environscitechnol}, and energy storage \cite{chen-2019-joule, yang-2025-sciadv}. Although preferential evaporation of the solvent is expected to gradually concentrate the solution until salt crystallization occurs, evaporation of saline solutions often gives rise to far more complex behavior \cite{lohse-2020-natrevphys, gelderblom-2022-sm}, including non-trivial flow patterns \cite{marin-2019-prf, bruning-2020-prf, mcbride-2024-acsami, martinezpuig-2026-jfm}, altered evaporation rates \cite{marin-2019-prf, bruning-2020-prf, seyfert-2022-prf, martinezpuig-2026-jfm}, and spatially heterogeneous deposition \cite{shahidzadeh-2008-langmuir, mailleur-2018-prl, mcbride-2018-langmuir, marin-2019-prf, mcbride-2021-sciadv, mcbride-2024-acsami}. \\

This complexity is further amplified under geometric confinement \cite{clement-2004-langmuir, rijniers-2005-prl, kohler-2018-prl, kohler-2022-natcommun}. Confined evaporation of salt solutions plays a critical role in technological applications, including carbon sequestration \cite{muller-2009-energyproc, miri-2015-intjgreenhgascontrol}, desalination membranes \cite{olufade-2018-acsomega}, inkjet printing \cite{rump-2023-prappl, rump-2025-prappl}, and microfluidic devices \cite{naillon-2017-jcrystgrowth}, while at the same time threatening natural ecosystems \cite{book-pitman, cooke-1981-procgeolassoc} and cultural heritage \cite{book-goudie, price-1997-archaeolint, charola-2000-jaminstconserv} through salt-induced damage. These considerations have driven renewed interest in quantitatively understanding evaporation dynamics in confined geometries such as porous media \cite{rijniers-2005-prl, steiger-2005-jcrystgrowth, sghaier-2009-transpporousmed, shahidzadeh-2010-pre, eloukabi-2011-chemengtechnol, schiro-2012-prl, verantissoires-2012-prl, naillon-2018-prl, wijnhorst-2024-prappl, luckins-2025-jfm, wijnhorst-2025-prappl, wijnhorst-2025-arxiv, prat-2026-langmuir} and single capillaries \cite{olbricht-1996-arfm, prat-2002-chemengj, chauvet-2009-prl, desarnaud-2014-jpcl, naillon-2015-jcrystgrowth, roger-2021-pnas, pingulkar-2023-sm, raju-2024-jfm, huisman-2025-sm}. Capillaries, in particular, offer a minimal and well-controlled platform \cite{raju-2024-jfm}: they enable precise control over unidirectional evaporation, allow optical access for real-time measurements, and isolate the key physical mechanisms that govern transport and phase change. \\

A striking manifestation of unidirectional evaporation in saline systems is \emph{salt creeping}, in which crystals precipitate and propagate away from the evaporating interface along surrounding solid surfaces \cite{book-buckley, washburn-1926-jphyschem, druce-1927-jpharm, erlenmeyer-1927-helvchemacta, hazlehurst-1936-jphyschem, huang-1976-nature, vanenckevort-2013-crystgrowthdes, qazi-2019-sciadv, wijnhorst-2025-arxiv, prat-2026-langmuir}. First reported nearly a century ago \cite{washburn-1926-jphyschem}, salt creeping has drawn attention ever since, both for its visually dramatic appearance and its practical consequences \cite{wijnhorst-2025-arxiv}. Despite early phenomenological descriptions \cite{washburn-1926-jphyschem, hazlehurst-1936-jphyschem, huang-1976-nature}, a unified physical understanding of the mechanisms governing salt creeping has remained elusive. Recent years have seen renewed interest in this phenomenon \cite{vanenckevort-2013-crystgrowthdes, qazi-2019-sciadv, wijnhorst-2025-arxiv}, motivated in part by its relevance to the energy transition \cite{miri-2015-intjgreenhgascontrol, kallenberger-2018-communchem}. \\

The prevailing qualitative picture posits that evaporation-induced crystallization builds an interconnected porous salt network, and that network wicks liquid up from the reservoir by capillary action \cite{vanenckevort-2013-crystgrowthdes, qazi-2019-sciadv}. Evaporation of this imbibed liquid deposits additional salt, extending the porous structure and reinforcing liquid transport, which feeds a self-amplifying creeping process \cite{desarnaud-2015-japplphys, miri-2015-intjgreenhgascontrol, qazi-2019-sciadv, wijnhorst-2025-arxiv, prat-2026-langmuir}. However, key mechanistic questions remain unresolved. In particular, how salt concentration and ambient relative humidity regulate evaporation dynamics, trigger phase transitions, and control the onset and persistence of salt creeping is still poorly understood. Addressing these questions is essential for predicting, controlling, and mitigating salt-induced damage, as well as exploiting evaporation of saline solutions and salt precipitation in technological applications. \\

A recurring theme across these confined systems is that steady external driving can produce distinctly unsteady dynamics. This is the signature of relaxation oscillators \cite{book-strogatz, book-fowler, book-dibernardo} --- systems in which slow loading alternates with rapid discharge across a threshold --- which appear throughout the physical sciences, from the dripping faucet \cite{book-shaw} and stick-slip friction \cite{watanabe-2025-pre} to geyser eruptions \cite{hurwitz-2017-arearthplanetsci}, electronic circuits \cite{vdpol-1926-philosmag, ginoux-2012-chaos}, and the saline density oscillator \cite{martin-1970-geophysfluiddyn, yoshikawa-1989-jchemeduc}. As we show below, confined salt creeping realizes precisely such an oscillator, with the threshold supplied by the supersaturation required to nucleate new crystals and the reset supplied by their dissolution at saturation --- that is, by the metastable zone of the salt itself \cite{desarnaud-2014-jpcl}. \\

Here we investigate the unidirectional evaporation of aqueous salt solutions under controlled confinement, focusing on the coupled roles of salt concentration and relative humidity in setting evaporation dynamics. Using a round cylindrical glass capillary geometry, we impose evaporation exclusively from one open end where the evaporating liquid meniscus is maintained. By systematically varying the initial salt concentration and ambient humidity, we show that evaporation dynamics shifts from smooth and monotonic to strongly intermittent as salt creeping develops on the outer capillary surface. We find that salt creeping introduces an additional pathway for water loss through imbibition into the growing crystal network, resulting in burst-like, intermittent evaporation dynamics. \\

To rationalize these observations, we develop a minimal theoretical model that captures the essential feedback between evaporation, imbibition, and phase change. The model reveals that the metastable zone of the salt --- the gap between the supersaturation at which crystals nucleate and the saturation concentration at which they redissolve --- introduces an intrinsic timescale into the system, and that the emergence and disappearance of intermittency with humidity and salt concentration follows from two conditions on the salt concentration at the capillary exit: the exit concentration must reach the crystallization supersaturation within the accessible time to precipitate crystals, and subsequently fall back below the saturation concentration to partially dissolve the precipitated crystals. Taken together, our results identify precipitation-dissolution cycling as a generic physical mechanism underlying intermittent salt creeping, and provide a quantitative framework for evaporation-driven phase transitions in confined saline systems. 

\section{Results}

\subsection{Unidirectional evaporation induces intermittent salt creeping}

\begin{figure}
    \centering
    \includegraphics[width=\textwidth]{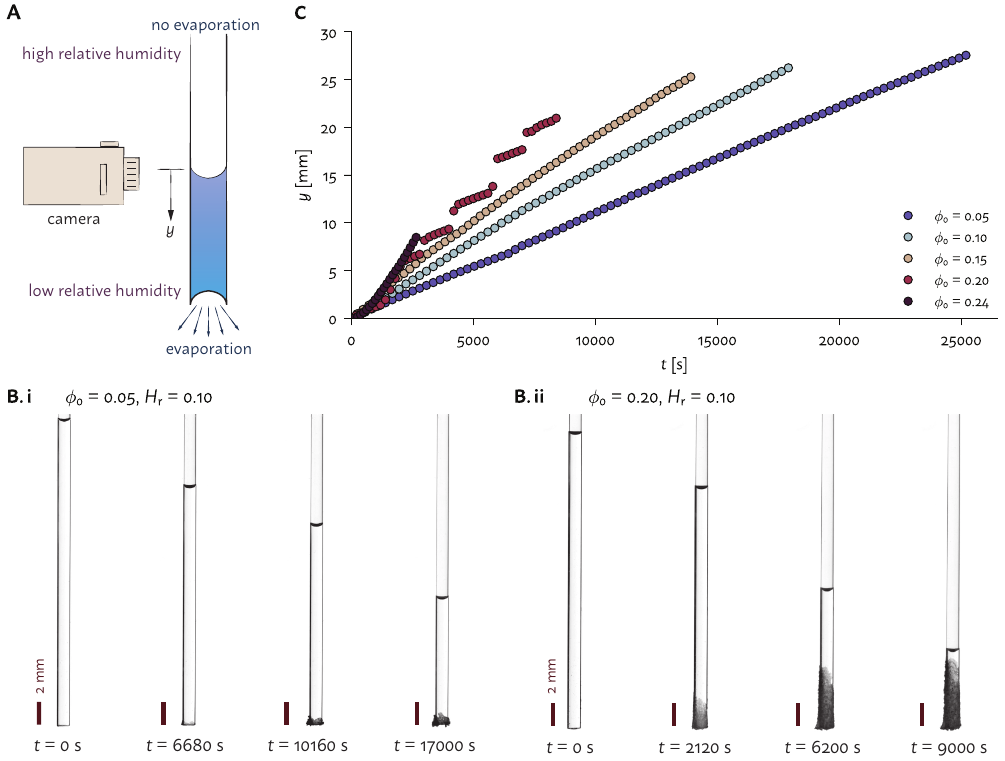}
    \caption{\textbf{Salt creeping driven by unidirectional evaporation.} \textbf{A.} Schematic of the experimental setup. Time-lapsed experimental snapshots at relative humidity $H_{\mathrm{r}}$ = 0.10 for initial salt concentration \textbf{B.i.} $\phi_{0}$ = 0.05, for which salt creeping is largely suppressed, and \textbf{B.ii.} $\phi_{0}$ = 0.20, showing pronounced salt creeping. \textbf{C.} Temporal evolution of the top-meniscus displacement $y (t)$ for different initial salt concentrations, $\phi_{0}$, at $H_{\mathrm{r}}$ = 0.10. Only a selection of experimental datapoints are shown to avoid overcrowding of the plot. See movies SM1, SM2, and SM3 for the corresponding experimental movies.}
    \label{fig:fig-1}
\end{figure}

To probe the role of dissolved salts in confined evaporation, we study aqueous sodium chloride (\textsf{NaCl}) solutions, of initial mass fraction $\phi_{0}$, evaporating unidirectionally from a cylindrical glass capillary maintained at a controlled relative humidity $H_{\mathrm{r}}$. Evaporation occurs exclusively at the lower pinned meniscus --- which remains spatially anchored at the bottom open end of the capillary --- causing the liquid column to shorten and the upper meniscus to recede downward. The temporal evolution of this recession, $y (t)$, is recorded by side-view imaging (figure \ref{fig:fig-1}A; see Materials and Methods for experimental details). \\

At low relative humidity ($H_{\mathrm{r}}$ = 0.10) and low initial salt mass fraction ($\phi_{0}$ = 0.05), the top meniscus recedes smoothly and monotonically with time (figure \ref{fig:fig-1}B.i and movie SM1). Quantitative measurements of $y (t)$ reveal an approximately linear temporal dependence (figure \ref{fig:fig-1}C), consistent with steady evaporation \cite{book-crank, raju-2024-jfm}. At the same time, salt precipitates near the lower opening of the capillary, forming a deposit on the outer surface that thickens gradually in the \emph{radial} direction as evaporation proceeds (figure \ref{fig:fig-1}B.i and movie SM1; see figure \ref{fig:fig-3} for a detailed discussion on directionality in crystal deposition). Similar smooth, near-linear meniscus dynamics are observed for intermediate salt concentrations ($\phi_{0}$ = 0.10 and 0.15), albeit with increased recession rates, as reflected by the higher slopes of $y (t)$ in figure \ref{fig:fig-1}C. \\

A qualitatively different behavior emerges at a higher salt concentration. For $\phi_{0}$ = 0.20 at the same relative humidity ($H_{\mathrm{r}}$ = 0.10), the top meniscus continues to recede but does so intermittently, through discrete step-like events separated by periods of gradual recession (figures \ref{fig:fig-1}B.ii and \ref{fig:fig-1}C, and movie SM2). This intermittent dynamics contrasts sharply with the smooth recession observed at lower $\phi_{0}$. In this regime, salt precipitation no longer leads to a progressive thickening of the deposit near the capillary mouth. Instead, the crystalline material spreads on the outer surface of the capillary along its length in the \emph{axial} direction, displaying the characteristic features of salt creeping \cite{qazi-2019-sciadv} (figure \ref{fig:fig-1}B.ii and movie SM2; see figure \ref{fig:fig-3} for a detailed description of directionality in crystal deposition). \\

Upon further increasing the initial salt concentration to $\phi_{0}$ = 0.24, close to the saturation limit of sodium chloride in water ($\phi_{\mathrm{sat}}$ = 0.267), salt creeping along the outer capillary surface persists (movie SM3). However, the intermittent top-meniscus dynamics is no longer observed. Instead, the meniscus recession accelerates in time, giving rise to a strongly nonlinear $y (t)$ dependence (figure \ref{fig:fig-1}C) --- reminiscent of the self-amplifying creeping behavior reported previously in literature \cite{miri-2015-intjgreenhgascontrol, qazi-2019-sciadv, wijnhorst-2025-arxiv}. These observations establish unidirectional evaporation as a driver of salt creeping and reveal a concentration-dependent sequence of three dynamical regimes at fixed humidity (figure \ref{fig:fig-1}C): smooth meniscus recession without sustained salt creeping ($\phi_{0}$ = 0.05), intermittent-step like recession ($\phi_{0}$ = 0.20), and continuous self-amplified growth ($\phi_{0}$ = 0.24). Self-amplified creeping is one of these three; what the present measurement adds is the identification of the conditions under which the other two occur instead.  

\subsection{Ambient humidity shifts recession dynamics independent of salt concentration}

\begin{figure}
    \centering
    \includegraphics[width=\textwidth]{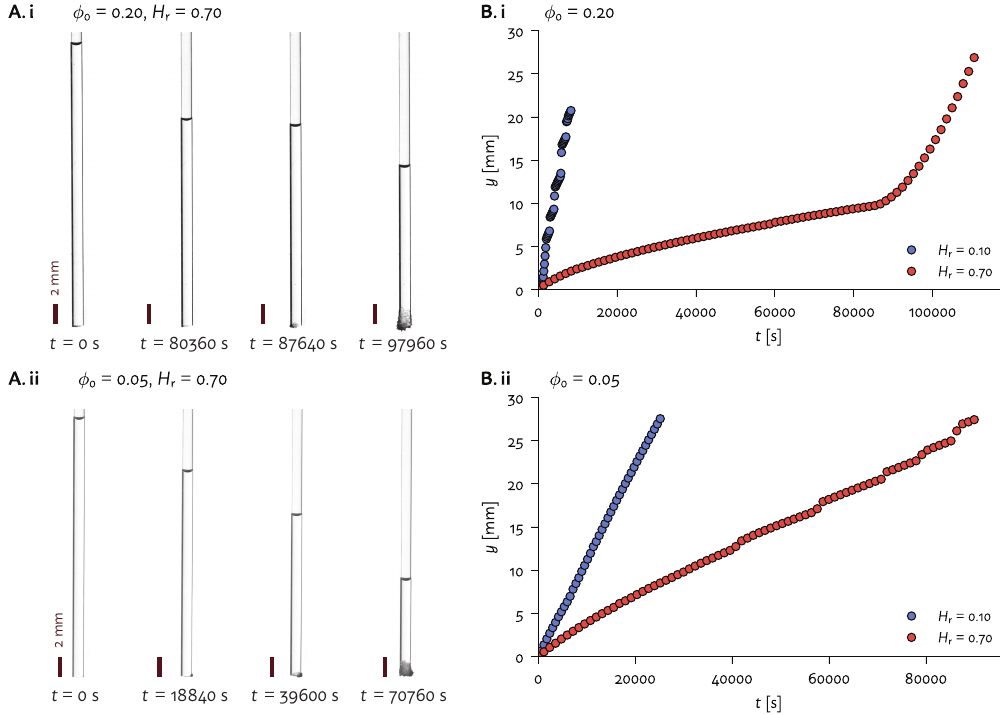}
    \caption{\textbf{Relative humidity can control both salt creeping and intermittent dynamics.} At $H_{\mathrm{r}}$ = 0.70, \textbf{A.i.} time-lapsed experimental snapshots at $\phi_{0}$ = 0.20 show salt creeping while \textbf{B.i.} the corresponding top-meniscus dynamics remain non-intermittent. In contrast, at $\phi_{0}$ = 0.05, \textbf{A.ii.} time-lapsed experimental snapshots reveal the onset of salt creeping, and \textbf{B.ii.} the top-meniscus dynamics display pronounced intermittency. Only a selection of experimental datapoints are shown to avoid overcrowding of the plots. See movies SM4 and SM5 for the corresponding experimental movies.}
    \label{fig:fig-2}
\end{figure}

In addition to the initial salt mass fraction $\phi_{0}$, the relative humidity $H_{\mathrm{r}}$ provides an independent control parameter by setting the rate of solvent loss and thereby the strength of the non-equilibrium driving. To isolate its role, we fix the initial salt mass fraction at $\phi_{0}$ = 0.20 and increase the relative humidity at the evaporating end to $H_{\mathrm{r}}$ = 0.70. The reduced vapor-pressure contrast between the liquid surface and the ambient leads to a substantially slower recession of the top meniscus (figure \ref{fig:fig-2}A.i and movie SM4), along with a delayed onset of crystallization and salt deposition on the outer surface of the capillary. \\

At this elevated humidity, the intermittent meniscus dynamics observed at lower $H_{\mathrm{r}}$ is suppressed. Instead, the top meniscus recedes smoothly in time, with $y (t)$ exhibiting a continuous temporal evolution (figure \ref{fig:fig-2}B.i). This smooth dynamics persists until the first crystalline deposit appears on the outer surface ($t \approx$ 87640 \SI{}{\text{\second}}; figure \ref{fig:fig-2}A.i), at which point the recession rate increases rapidly (figure \ref{fig:fig-2}B.i). Following nucleation, salt deposition proceeds continuously and displays clear axial salt-creeping behavior along the outer capillary surface (figure \ref{fig:fig-2}A.i and movie SM4). Notably, even with the presence of salt creeping, the meniscus dynamics remains non-intermittent, showing strongly nonlinear yet continuous temporal evolution --- resembling the self-amplifying creeping dynamics reported previously in literature \cite{miri-2015-intjgreenhgascontrol, qazi-2019-sciadv, wijnhorst-2025-arxiv} (figure \ref{fig:fig-2}B.i). \\

A contrasting response is observed when the initial salt concentration is reduced while maintaining the same elevated humidity. For $\phi_{0}$ = 0.05 at $H_{\mathrm{r}}$ = 0.70, the recession rate of the top meniscus is again reduced relative to the low-humidity case (figures \ref{fig:fig-2}A.ii and \ref{fig:fig-2}B.ii, and movie SM5). However, in contrast to the smooth dynamics observed at higher $\phi_{0}$, the meniscus motion now exhibits intermittent step-like jumps (figure \ref{fig:fig-2}B.ii). Although these events are less pronounced than those observed at $H_{\mathrm{r}}$ = 0.10 and $\phi_{0}$ = 0.20 (figure \ref{fig:fig-2}B.i), their presence signals a re-emergence of intermittency. At the same time, the salt deposit develops into an axial creeping morphology along the outer surface of the capillary (figure \ref{fig:fig-2}A.ii and movie SM5), in stark contrast to the purely radial deposition observed for the same $\phi_{0}$ at a lower humidity of $H_{\mathrm{r}}$ = 0.10 (figure \ref{fig:fig-1}B.i and movie SM1). \\

Together, these observations show that salt concentration and relative humidity jointly regulate both salt creeping and intermittent top-meniscus dynamics. In addition to these two parameters, the nucleation rate of salt crystals during the precipitation phase constitutes a third control parameter influencing the top-meniscus dynamics \cite{townsend-2016-crystengcomm, qazi-2017-langmuir, qazi-2019-sciadv}, as demonstrated in figures \ref{fig:fig-3}D and \ref{fig:fig-s2}, and movie SM6. We return to nucleation control before the theoretical model has been introduced. In what follows, we focus on the physical mechanisms behind intermittent dynamics during confined, unidirectional evaporation of salt solutions. 

\subsection{Imbibition-mediated feedback facilitates intermittent salt creeping}

\begin{figure}
    \centering
    \includegraphics[width=\textwidth]{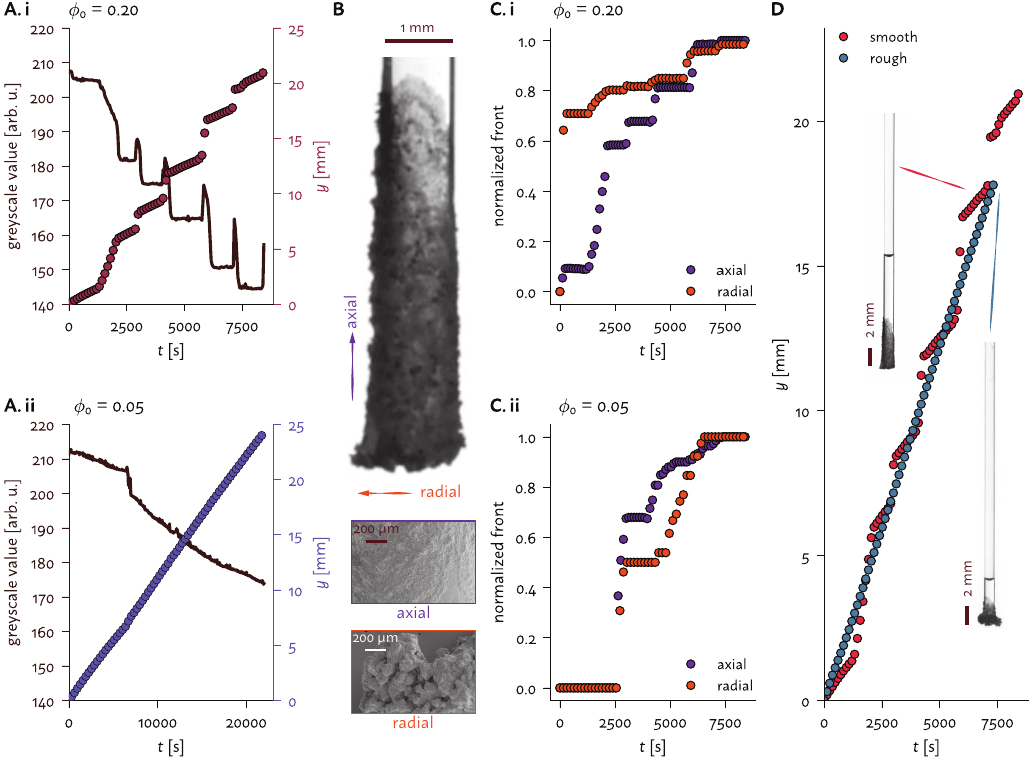}
    \caption{\textbf{Imbibition-mediated feedback and its morphological and nucleation control.} At $H_{\mathrm{r}}$ = 0.10 and \textbf{A.i.} $\phi_{0}$ = 0.20, time-resolved measurements of the local greyscale intensity (solid line; left axis) exhibit pronounced transient dips that flag episodic water imbibition into the creeping salt; these events are temporally correlated with the step-like jumps in the top meniscus position $y (t)$ (discrete markers; right axis). By comparison, \textbf{A.ii.} for $\phi_{0}$ = 0.05, the greyscale intensity evolves smoothly in time, consistent with the absence of salt creeping and continuous top-meniscus motion (the small dip at $t \approx$ 6680 \SI{}{\text{\second}} corresponds to the appearance of the first salt crystals). \textbf{B.} Salt creeping proceeds via coupled axial and radial growth: representative experimental snapshot illustrating axial and radial growth modes. Scanning electron micrographs reveal the contrasting microstructures: axial growth corresponds to small, densely interspersed crystals while radial growth consists of loosely packed, larger crystalline structures. \textbf{C.i.} At $\phi_{0}$ = 0.20 the axial front advances through discrete step-like jumps while the radial front grows gradually, whereas \textbf{C.ii.} at $\phi_{0}$ = 0.05 both fronts advance in tandem. \textbf{D.} Promoting crystal nucleation quenches creeping and intermittency: on a smooth capillary the meniscus recedes intermittently, whereas roughening the capillary base to enhance nucleation yields smooth, continuous recession; snapshots in the insets show the corresponding deposit morphologies. Only a selection of experimental datapoints are shown to avoid overcrowding of the plots. See movies SM2 and SM6 for the corresponding experimental movies. }
    \label{fig:fig-3}
\end{figure}

The intermittent top-meniscus dynamics observed at low relative humidity ($H_{\mathrm{r}}$ = 0.10) and high initial salt concentration ($\phi_{0}$ = 0.20) is accompanied by pronounced fluctuations in the optical appearance of the salt deposit (figure \ref{fig:fig-1}B.ii and movie SM2). This behavior is quantified by tracking the temporal evolution of the greyscale intensity of the crystal deposit-covered region in the experimental images (figure \ref{fig:fig-3}A.i), where values range from 0 (black) to 255 (white). Although the greyscale intensity decreases overall with time --- reflecting the progressive coverage of the capillary surface by crystalline salt --- it also exhibits distinct, burst-like events. \\

These bursts consist of a rapid increase in greyscale intensity followed by a pronounced dip. The deposited crystals thus form a permeable network into which salt solution from the capillary imbibes \cite{sghaier-2009-transpporousmed, eloukabi-2011-chemengtechnol, desarnaud-2015-japplphys, qazi-2019-sciadv, lazhar-2020-pof, wijnhorst-2024-prappl, wijnhorst-2025-arxiv}. The intensity increase corresponds to a change in refractive index associated with liquid infiltration into the porous salt deposit. Evaporation of this imbibed liquid at the outer surface then precipitates additional salt, extending the porous network further along the capillary and enabling yet more imbibition --- this is reflected in the sharp dip of the greyscale value. The additional deposited salt makes the greyscale value to fall below its pre-burst value. Salt creeping is therefore a process sustained by a feedback between evaporation and imbibition --- a mechanism reported in a variety of geometries where salt creeping is observed \cite{miri-2015-intjgreenhgascontrol, qazi-2019-sciadv, licsandru-2019-pre, lazhar-2020-pof, wijnhorst-2025-arxiv}. \\

Strikingly, the burst-like events in greyscale intensity are temporally correlated with the intermittent, step-like jumps in the top-meniscus position (figure \ref{fig:fig-3}A.i). This correlation pins down imbibition-mediated feedback as the physical origin of the intermittent meniscus dynamics observed under these conditions. \\

By contrast, when salt creeping is absent ($H_{\mathrm{r}}$ = 0.10 and $\phi_{0}$ = 0.05; figure \ref{fig:fig-1}B.i and movie SM1), the greyscale intensity decreases smoothly in time without burst-like fluctuations (figure \ref{fig:fig-3}A.ii). Apart from a single, weak dip in the greyscale intensity coinciding with the appearance of the first crystalline deposit ($t \approx$ 6680 \SI{}{\text{\second}}; figure \ref{fig:fig-1}B.i and movie SM1), the evolution remains gradual. This dip aligns temporally with a minor kink in the otherwise smooth top-meniscus dynamics ($t \approx$ 6680 \SI{}{\text{\second}}; figure \ref{fig:fig-3}A.ii), which points to the absence of sustained imbibition-mediated feedback in this regime. The lack of such feedback explains both the suppression of salt creeping and the continuous, non-intermittent meniscus motion. \\

The imbibition feedback also imprints a characteristic signature on the \emph{morphology} of the growing deposit. Tracking the axial and radial crystal fronts (figure \ref{fig:fig-3}B) shows that, at $H_{\mathrm{r}}$ = 0.10 and $\phi_{0}$ = 0.20, the radial front advances smoothly while the axial front progresses through discrete, step-like jumps (figure \ref{fig:fig-3}C.i) --- the morphological counterpart of the intermittent meniscus dynamics. Scanning electron micrographs (figure \ref{fig:fig-3}B) reveal the underlying microstructural contrast: the axially propagating deposit consists of small, densely packed crystals in a quasi-planar arrangement, whereas the radial deposit is composed of larger, loosely packed crystals forming a highly porous structure. At the same humidity but lower concentration ($\phi_{0}$ = 0.05), both fronts advance in tandem (figure \ref{fig:fig-3}C.ii), consistent with the absence of sustained creeping. The two fronts differ in growth mode as well as in rate, and the micrographs record that difference. For sodium chloride, crystal habit, size, and packing are known to depend on how fast the solution is driven --- rapid growth producing many small, densely packed crystals and slower growth fewer, larger, and more openly packed ones \cite{shahidzadeh-2008-langmuir, verantissoires-2012-prl, desarnaud-2014-jpcl, wijnhorst-2025-prappl}. The axial front, which advances in discrete episodes (figure \ref{fig:fig-3}C.i), builds the small, densely interspersed crystals of the micrograph; the radial front, which advances continuously at a lower rate, builds the larger and more open structure. The microstructural contrast of figure \ref{fig:fig-3}B and the kinematic contrast of figure \ref{fig:fig-3}C are therefore two records of the same difference.  \\

Finally, the feedback can be switched off by controlling where crystals nucleate (figure \ref{fig:fig-3}D). On a smooth (as-received) capillary the deposit creeps axially with intermittent meniscus dynamics (for $\phi_{0}$ = 0.20 and $H_{\mathrm{r}}$ = 0.10). When the glass at the capillary exit is roughened to enhance nucleation, axial creeping and its associated intermittency are suppressed; the deposit grows predominantly radially, and the meniscus recedes smoothly (figure \ref{fig:fig-3}D and movie SM6). We note that roughening a hydrophilic surface would, on wetting grounds alone \cite{book-degennes}, be expected to promote the spreading of the solution rather than to suppress creeping. What the experiment shows is that spreading is not suppressed --- the deposit grows abundantly --- but that its \emph{axial propagation} is. The material accumulates radially at the mouth instead. This points to the ease of nucleation, rather than the wettability of the wall, as the variable that roughening controls. The theoretical model in the following section, extended to include nucleation kinetics in figure \ref{fig:fig-s2}, shows that increasing the density of nucleation sites reproduces the observed suppression.   \\

Put together, these results point to evaporation-imbibition feedback as the immediate physical mechanism driving intermittent salt creeping and meniscus dynamics. However, the emergence and persistence of this feedback require a mechanism that periodically activates and deactivates imbibition during evaporation. In the following section, we show that the metastable zone of the salt \cite{desarnaud-2014-jpcl} --- the concentration interval between the supersaturation at which crystals nucleate and the saturation concentration at which they redissolve --- provides precisely such a mechanism, giving rise to the observed intermittency. Mechanisms other than a precipitation-dissolution cycle that could in principle produce intermittency here --- solutal Marangoni flow \cite{marin-2019-prf}, osmotic or ionic effects at the charged crystal surfaces \cite{book-israelachvili}, Liesegang-type periodic precipitation \cite{nabika-2020-langmuir}, and stick-slip depinning of the evaporating interface \cite{watanabe-2025-pre} --- are set against our measurements in the Supplementary Information, and are found to be inconsistent with them. 

\subsection{A crystallization relaxation oscillator}

\begin{figure}
    \centering
    \includegraphics[width=\textwidth]{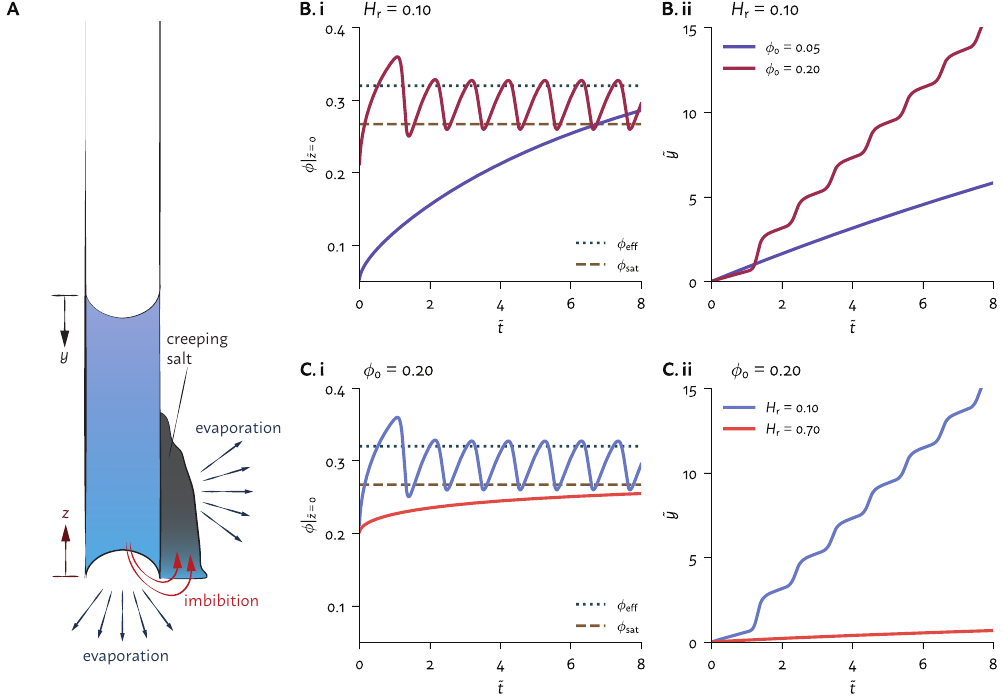}
    \caption{\textbf{Precipitation-dissolution cycling governs intermittent dynamics.} \textbf{A.} Schematic of the geometry for the theoretical model, showing both the evaporation and imbibition pathways of mass loss from the capillary. Temporal evolution of \textbf{B.i.}, \textbf{C.i.} the salt concentration at the capillary exit, $\phi \vert_{\tilde{z} = 0}$, and \textbf{B.ii.}, \textbf{C.ii.} the top-meniscus position, $\tilde{y}$, shown for $\phi_{0}$ = 0.05 and 0.20 at $H_{\mathrm{r}}$ = 0.10, and for $H_{\mathrm{r}}$ = 0.10 and 0.70 at $\phi_{0}$ = 0.20, respectively. Crystallization begins once $\phi \vert_{\tilde{z} = 0}$ exceeds the efflorescence threshold $\phi_{\mathrm{eff}}$ (dotted lines in panels B.i and C.i) and the deposit dissolves once it falls below the saturation mass fraction $\phi_{\mathrm{sat}}$ (dashed lines in panels B.i and C.i), so that the interval between the two thresholds --- the metastable zone of the salt --- closes a cycle of slow loading and fast discharge that operates entirely in the liquid phase. Oscillations of $\phi \vert_{\tilde{z} = 0}$ between $\phi_{\mathrm{eff}}$ and $\phi_{\mathrm{sat}}$ coincide with intermittent meniscus dynamics, whereas the absence of such oscillations yields smooth temporal evolution.}
    \label{fig:fig-4}
\end{figure}

To quantitatively rationalize how the initial salt concentration $\phi_{0}$ and the ambient relative humidity $H_{\mathrm{r}}$ determine whether the top-meniscus dynamics $y(t)$ is intermittent or monotonic (figure \ref{fig:fig-1}C), we develop a minimal theoretical model that captures the dominant modes of water mass loss from the capillary and accounts for the evaporation-imbibition feedback. The salt solution inside the capillary, of radius $R$, is modeled as a one-dimensional, isothermal, semi-infinite liquid column, with the lower liquid-air interface located at the capillary exit ($z$ = 0) and exposed to a fixed relative humidity $H_{\mathrm{r}}$ (figure \ref{fig:fig-4}A). This framework has been shown to successfully describe unidirectional evaporation dynamics of multicomponent liquids under confinement \cite{raju-2024-jfm}. \\

The local composition of the solution is described by the salt mass fraction $\phi (z, t)$, with the corresponding water mass fraction given by $1 - \phi(z,t)$. The spatiotemporal evolution of $\phi$ is governed by the one-dimensional advection-diffusion equation, derived under the assumption of constant solution density $\rho$ and diffusivity $\mathcal{D}$ corresponding to the initial composition $\phi_{0}$, given by \cite{raju-2024-jfm}
\begin{align}
    \frac{\partial \phi}{\partial t} + u \, \frac{\partial \phi}{\partial z} = \mathcal{D} \frac{\partial^{2} \phi}{\partial z^{2}} ,
    \label{eq:adv-diff}
\end{align}
where $u(t)$ denotes the advection velocity (the purely temporal dependence follows from the one-dimensional continuity equation under the constant-density assumption \cite{raju-2024-jfm}). Initially, the salt concentration is uniform throughout the liquid column and equal to $\phi_{0}$. \\

Water is lost exclusively at the lower interface at $z$ = 0, which is exposed to the ambient relative humidity $H_{\mathrm{r}}$, while evaporation from the upper interface is negligible \cite{raju-2024-jfm}. This water loss at the lower interface is replenished by advective and diffusive transport from the bulk liquid, giving rise to a boundary condition at $z$ = 0 involving the total water flux $J (t)$, expressed as
\begin{align}
    - u \left( 1 - \phi \vert_{z = 0} \right) + \mathcal{D} \, \frac{\partial \left( 1 - \phi \right)}{\partial z} \bigg \vert_{z = 0} = \frac{J}{\rho} .
    \label{eq:water-conserv}
\end{align}
In contrast to typical models of unidirectional evaporation of multicomponent liquids \cite{raju-2024-jfm}, this total flux does not arise solely from evaporation at the meniscus. In the present system, water is also extracted through imbibition into the salt crust that forms on the outer surface of the capillary during salt creeping (figure \ref{fig:fig-4}A). We therefore decompose \cite{lazhar-2020-pof} the total water flux $J (t)$ into an evaporative contribution from the meniscus, $J_{\mathrm{meniscus}} (t)$, and an imbibition or wicking contribution into the salt deposit, $J_{\mathrm{wick}} (t)$, i.e. $J = J_{\mathrm{meniscus}} + J_{\mathrm{wick}}$. That liquid transport through a porous salt structure can carry water at a rate comparable to or exceeding that supplied by evaporation from the meniscus alone has recently been established independently, in a theoretical treatment of salt structures growing on a porous substrate \cite{prat-2026-langmuir}. The evaporative flux is estimated using a quasi-steady, diffusion-limited description of evaporation from a pinned interface \cite{popov-2005-pre, seyfert-2022-prf, gelderblom-2022-sm, wilson-2023-arfm, raju-2024-jfm, martinezpuig-2026-jfm}, given by
\begin{align}
    J_{\mathrm{meniscus}} = \frac{4 \mathcal{D}_{\mathrm{v}}}{\pi R} c_{\mathrm{sat}} \left( \chi - H_{\mathrm{r}} \right),
    \label{eq:j-meniscus}
\end{align}
where $c_{\mathrm{sat}}$ is the saturation mass concentration of water vapor at the surface of pure water, $\mathcal{D}_{\mathrm{v}}$ the diffusion coefficient of water vapor in air (assumed to be independent of the concentration of water vapor \cite{raju-2024-jfm}), and $\chi$ the thermodynamic activity of water corresponding to the concentration \cite{mikhailov-2004-atmoschemphys, roger-2021-pnas, seyfert-2022-prf, raju-2024-jfm, martinezpuig-2026-jfm} $\phi \vert_{z = 0}$ at $z = 0$. The salt rejected at the evaporating interface is assumed to be carried entirely into the wicking flux, yielding a coupled boundary condition for the salt concentration at $z = 0$:
\begin{align}
    - u \, \phi \vert_{z = 0} + \mathcal{D} \, \frac{\partial \phi}{\partial z} \bigg \vert_{z = 0} = \frac{J_{\mathrm{wick}}}{\rho} \left( \frac{\phi \vert_{z = 0}}{1 - \phi \vert_{z = 0}} \right).
    \label{eq:salt-conserv}
\end{align}

The liquid imbibing into the salt crust has concentration $\phi \vert_{z = 0}$ and is itself exposed to the same ambient humidity $H_{\mathrm{r}}$. Evaporation from the porous salt deposit is therefore modeled using Fickian vapor diffusion in lumped form \cite{book-crank, book-levich}, characterized by an effective diffusion length $\delta_{\mathrm{v}}$ and a temporally varying wicking surface area $A_{\mathrm{wick}} (t)$, given by
\begin{align}
    J_{\mathrm{wick}} = \frac{\mathcal{D}_{\mathrm{v}} c_{\mathrm{sat}}}{\pi R^{2} \delta_{\mathrm{v}}} A_{\mathrm{wick}} \left( \chi - H_{\mathrm{r}} \right).
    \label{eq:j-wick}
\end{align}
Importantly, $A_{\mathrm{wick}}$ denotes the area of the porous crystalline network that actively participates in wicking and evaporation, rather than the total external crystal surface area. However, both these areas are of the same order of magnitude. \\

At this point, the solution of concentration $\phi \vert_{z = 0}$ entering the salt wick either deposits additional salt or dissolves existing crystals, and which occurs depends on how $\phi \vert_{z = 0}$ compares with two distinct concentrations. The lower is the saturation mass fraction $\phi_{\mathrm{sat}}$: below it the wicked solution is unsaturated with respect to the deposit and dissolves it. The upper is the efflorescence threshold $\phi_{\mathrm{eff}}$: crystallization does not begin as soon as the solution becomes supersaturated but requires a finite supersaturation for crystals to nucleate, so that $\phi_{\mathrm{eff}} > \phi_{\mathrm{sat}}$. The interval between them is the metastable zone of the salt, which for sodium chloride in confinement has been measured directly \cite{desarnaud-2014-jpcl}.  \\

Evaporation concentrates the exit until $\phi \vert_{z = 0}$ reaches $\phi_{\mathrm{eff}}$. Crystals then nucleate on the existing deposit and $A_{\mathrm{wick}}$ grows. The enlarged wick draws liquid out of the capillary faster than evaporation from the meniscus alone would, so the exit is simultaneously flushed with less concentrated solution from the bulk and depleted by the precipitation itself, and $\phi \vert_{z = 0}$ falls. Once it drops below $\phi_{\mathrm{sat}}$ the wicked solution is unsaturated, which dissolves the finest asperities of the salt deposit, and $A_{\mathrm{wick}}$ shrinks, whereupon evaporation reconcentrates the exit. We capture these processes using first-order kinetics for both precipitation and dissolution, characterized by a single material parameter $\delta_{\mathrm{eff}}$ or $\delta_{\mathrm{sat}}$, and a two-state (growth/dissolution) switch $\mathrm{s}$ that encodes the hysteresis:
\begin{align}
    \frac{\mathrm{d} A_{\mathrm{wick}}}{\mathrm{d}t} =& \frac{J_{\mathrm{wick}}}{\rho \, \delta_{\mathrm{s}}} \, A_{\mathrm{wick}} \left( \frac{\phi \vert_{z = 0}}{1 - \phi \vert_{z = 0}} \right) \left( \phi \vert_{z = 0} - \phi_{\mathrm{s}} \right), \nonumber \\ & \mathrm{where}
    \begin{cases}
       \mathrm{s} = \mathrm{growth}, \, \delta_{\mathrm{s}} = \delta_{\mathrm{eff}}, \, \phi_{\mathrm{s}} = \phi_{\mathrm{sat}} & \mathrm{after} \, \, \phi \vert_{z = 0} \, \, \mathrm{last \, \, reached} \, \, \phi_{\mathrm{eff}}, \\
       \mathrm{s} = \mathrm{dissolution}, \, \delta_{\mathrm{s}} = \delta_{\mathrm{sat}}, \, \phi_{\mathrm{s}} = \phi_{\mathrm{eff}} & \mathrm{after} \, \, \phi \vert_{z = 0} \, \, \mathrm{last \, \, fell \, \, below} \, \, \phi_{\mathrm{sat}}.
    \end{cases}
    \label{eq:hysteresis}
\end{align}
Equations \eqref{eq:j-wick} and \eqref{eq:hysteresis} close a cycle marked by slow, transport-limited loading, a threshold-triggered fast discharge, and a hysteretic reset --- the defining structure of a relaxation oscillator \cite{book-strogatz,book-dibernardo}. We emphasize that this cycle operates entirely in the liquid phase: at the relative humidities studied, all well below the deliquescence relative humidity of sodium chloride ($\approx$ 0.75 at room temperature), \textsf{NaCl} crystals take up no water from ambient water vapor \cite{wise-2007-jgeophysres} and deliquescence cannot occur. The hysteresis invoked here is the one familiar from humidity-cycling experiments, expressed in a different variable --- there the metastable zone appears as a gap in humidity and the crystallization-dissolution cycle is imposed externally \cite{bellezza-2025-arxiv}, whereas here the humidity is constant and the cycle is self-generated. A salt crust migrating by dissolution at one face and precipitation at another has been described in evaporating porous media \cite{licsandru-2019-pre}; the present system is its confined, self-induced oscillating counterpart.   \\

Introducing characteristic length and time scales $\mathcal{L}_{\mathrm{D}} = \rho \pi \mathcal{D} R / 4 \mathcal{D}_{\mathrm{v}} \, c_{\mathrm{sat}}$ and $\mathcal{T}_{\mathrm{D}} = \mathcal{L}_{\mathrm{D}}^{2} / \mathcal{D}$ that arise from the competition between liquid-phase transport and vapor diffusion \cite{book-crank, raju-2024-jfm, martinezpuig-2026-jfm}, we non-dimensionalize the governing equations and boundary conditions (equations \eqref{eq:adv-diff} - \eqref{eq:hysteresis}, with tildes denoting non-dimensional quantities; see Materials and Methods), yielding a closed dynamical system coupling advection-diffusion in the liquid column to hysteretic evolution of the wicking surface area. Numerically solving this coupled dynamical system (see Supplementary Information for details) provides quantitative insight on the temporal variation of the salt concentration at the capillary exit, $\phi \vert_{\tilde{z} = 0}$, and the top-meniscus position, $\tilde{y} (\tilde{t})$. Because lengths and times are scaled with $\mathcal{L}_{\mathrm{D}}$ and $\mathcal{T}_{\mathrm{D}} = \mathcal{L}_{\mathrm{D}}^{2} / \mathcal{D}$, respectively, the velocity scale is $\mathcal{D} / \mathcal{L}_{\mathrm{D}}$, and the P\'{e}clet number of the exit region, $\mathrm{Pe} = \vert u \vert \, \mathcal{L}_{\mathrm{D}} / \mathcal{D}$, is identically the magnitude of the non-dimensional advection velocity $\tilde{u}$ (see Materials and Methods). It is therefore an output of the numerical solution rather than a quantity that has to be assumed, and is available at every instant for every condition (figure \ref{fig:fig-s3}).  \\

For $H_{\mathrm{r}}$ = 0.10 and $\phi_{0}$ = 0.20, the model predicts oscillations of the salt concentration at the capillary exit, $\phi \vert_{\tilde{z} = 0}$, between $\phi_{\mathrm{sat}}$ and $\phi_{\mathrm{eff}}$, as shown in figure \ref{fig:fig-4}B.i. These oscillations intermittently activate and suppress wicking, resulting in step-like, intermittent top-meniscus dynamics (figure \ref{fig:fig-4}B.ii), in close qualitative agreement with the experimental observations (figure \ref{fig:fig-1}C; see figure \ref{fig:fig-s1}A.i for a quantitative comparison). In contrast, for $H_{\mathrm{r}}$ = 0.10 and $\phi_{0}$ = 0.05, the concentration at the capillary exit, $\phi \vert_{\tilde{z} = 0}$, increases more slowly and crosses $\phi_{\mathrm{sat}}$ but not $\phi_{\mathrm{eff}}$ within the experimental time window (figure \ref{fig:fig-4}B.i), yielding smooth, monotonic dynamics (figure \ref{fig:fig-4}B.ii) --- also in accordance with the experimental observations (figure \ref{fig:fig-1}C). While the model suggests that $\phi \vert_{\tilde{z} = 0}$ may cross $\phi_{\mathrm{eff}}$ at much longer times for this parameter set (figure \ref{fig:fig-4}B.i), leading to oscillations, such behavior lies beyond the experimentally accessible regime in the present system and is not pursued further. \\

The model further captures the effect of ambient relative humidity $H_{\mathrm{r}}$. Increasing the relative humidity to $H_{\mathrm{r}}$ = 0.70 for $\phi_{0}$ = 0.20 significantly reduces the evaporation rate, thus slowing the temporal evolution of $\phi \vert_{\tilde{z} = 0}$ and suppressing oscillations between $\phi_{\mathrm{sat}}$ and $\phi_{\mathrm{eff}}$ (figure \ref{fig:fig-4}C.i). As a result, the predicted dynamics remains continuous and non-intermittent (figure \ref{fig:fig-4}C.ii), consistent with experimental observations (figure \ref{fig:fig-2}B.i). The solutions also show what the regimes correspond to in transport terms. $\mathrm{Pe}$ lies below unity throughout the loading intervals of the intermittent condition and rises by an order of magnitude at each crystallization event (figure \ref{fig:fig-s3}B) as $\tilde{A}_{\mathrm{wick}}$ jumps (figures \ref{fig:fig-s1}B.iii and \ref{fig:fig-s4}B), reaching 9.8 at $\phi_{0}$ = 0.20, so the exit alternates once per cycle between diffusion-controlled ($\mathrm{Pe} < 1$) loading and advection-controlled ($\mathrm{Pe} > 1$) discharge. The conditions that recede smoothly never leave the diffusive regime (figure \ref{fig:fig-s3}A).  \\

The quantitative coupling between the precipitation-dissolution switch and unidirectional advection-diffusion depends sensitively on the two threshold concentrations, $\phi_{\mathrm{eff}}$ and $\phi_{\mathrm{sat}}$, as well as on the characteristic microscopic length scales $\delta_{\mathrm{v}}$, $\delta_{\mathrm{eff}}$, and $\delta_{\mathrm{sat}}$. For sodium chloride, $\phi_{\mathrm{sat}}$ = 0.267 is a material constant, fixed by the solubility. In the present relay description (equation \eqref{eq:hysteresis}), $\phi_{\mathrm{eff}}$ (= 0.320) is the concentration at which a new crystallization event nucleates at the \emph{existing} creeping deposit --- heterogeneously assisted by the existing crust and walls, hence below the metastability limit for \emph{pristine} bulk nucleation in a clean capillary \cite{desarnaud-2014-jpcl}, but still a barrier-limited nucleation threshold rather than barrierless growth (figure \ref{fig:fig-s2}). Because it is nucleation-controlled, enhancing nucleation (roughened base; figure \ref{fig:fig-3}D and movie SM6) lowers $\phi_{\mathrm{eff}}$ toward $\phi_{\mathrm{sat}}$, collapsing the interval between the thresholds and suppressing the oscillations. Replacing the fixed threshold by nucleation kinetics, anchored to the independently measured metastability limit for sodium chloride in confinement \cite{desarnaud-2014-jpcl}, recovers the calibrated value of $\phi_{\mathrm{eff}}$ and shows how it shifts with drying rate and with the density of nucleation sites (figure \ref{fig:fig-s2}).   \\

The length scales $\delta_{\mathrm{v}}$, $\delta_{\mathrm{eff}}$, and $\delta_{\mathrm{sat}}$ arise from molecular-scale transport processes that are difficult to access experimentally, but their orders of magnitude can be reasonably estimated \cite{licsandru-2019-pre, lazhar-2020-pof}; these estimates guide the parameter choices used to predict the dynamics shown in figure \ref{fig:fig-4}. Every parameter entering the model is listed with its provenance in the Supplementary Information; of those entering the results reported here, two are calibrated and the remainder are literature values or set by the experimental design. One consequence of this minimal description is that the precise onset of crystallization is not predicted --- in alignment with the fact that precipitation kinetics is fast compared to transport kinetics \cite{naillon-2015-jcrystgrowth, mcbride-2021-sciadv}. Even so, the model still reproduces the experimentally observed smooth dynamics for $\phi_{0}$ = 0.05 and $H_{\mathrm{r}}$ = 0.10 (figure \ref{fig:fig-1}C) till the time when $\phi \vert_{\tilde{z} = 0}$ crosses $\phi_{\mathrm{eff}}$ (figures \ref{fig:fig-4}B.i and \ref{fig:fig-4}B.ii; see figure \ref{fig:fig-s1}A for a quantitative comparison), capturing the onset of creeping. With literature material parameters and no adjustment of the vapor-transport scale, the model reproduces the measured smooth-regime recession rate (figure \ref{fig:fig-s1}A.i), and --- for $\phi_{0}$ = 0.20, $H_{\mathrm{r}}$ = 0.10 --- the mean inter-event period and the individual step timings (figure \ref{fig:fig-1}C; see figure \ref{fig:fig-s1}B.i for a quantitative comparison). The model also predicts, qualitatively, the late-time self-amplified recession dynamics for $\phi_{0}$ = 0.20, $H_{\mathrm{r}}$ = 0.70 (figure \ref{fig:fig-2}B.i; see figure \ref{fig:fig-s1}C.i for a quantitative comparison).   \\

The model also resolves why the concentration trend reverses with humidity. Sustained cycling requires two conditions to hold together. The exit concentration must first reach $\phi_{\mathrm{eff}}$, so that a crystallization event is triggered; following the discharge it must then fall back below $\phi_{\mathrm{sat}}$, so that $A_{\mathrm{wick}}$ can shrink and a new loading interval can begin. Intermittency is the regime in which both conditions are satisfied repeatedly; each of the two non-intermittent regimes corresponds to one of them failing (figure \ref{fig:fig-s1}). Where the first fails the meniscus recedes smoothly, which is what the model predicts for $\phi_{0}$ = 0.05, $H_{\mathrm{r}}$ = 0.10 (figure \ref{fig:fig-4}B.i). Where the second is only marginally met, the deposit grows without effective reset. At $\phi_{0}$ = 0.24 the bulk reservoir is within 0.027 of saturation, so the solution resupplied to the exit is almost saturated. The exit concentration barely dips below $\phi_{\mathrm{sat}}$ after each discharge, reaching 0.264 against $\phi_{\mathrm{sat}}$ = 0.267 (figure \ref{fig:fig-s4}A), and the unsaturated interval is too short for $A_{\mathrm{wick}}$ to relax (figure \ref{fig:fig-s4}B). The wicking area therefore ratchets upward from cycle to cycle, ending the run some eightfold larger than at $\phi_{0}$ = 0.20, and the recession becomes effectively continuous --- manifesting the self-amplified behavior (figure \ref{fig:fig-1}C). Additionally, after the first cycle, the P\'{e}clet number $\mathrm{Pe}$ for the loading intervals no longer return below unity, and transport at the exit is advection-dominated at all times (figure \ref{fig:fig-s3}C). The margin available for reset shrinks steeply as $\phi_{0}$ approaches saturation --- to leading order the meniscus must recede by a fraction $(1 - \phi_{0} / \phi_{\mathrm{sat}})$ of the column for the bulk itself to saturate, which is $\approx$ 81\% at $\phi_{0}$ = 0.05, $\approx$ 25\% at $\phi_{0}$ = 0.20, and $\approx$ 10\% at $\phi_{0}$ = 0.24.   \\

Both conditions depend on humidity, which is what reverses the trend. Our measurements show that cumulative water loss is not the quantity that selects the regime: at $\phi_{0}$ = 0.05, the $H_{\mathrm{r}}$ = 0.10 and $H_{\mathrm{r}}$ = 0.70 cases reach the same meniscus displacement, $y \approx$ 27.6 \SI{}{\text{\milli\meter}}, over $\approx$ 25240 \SI{}{\text{\second}} and 90680 \SI{}{\text{\second}}, respectively (figure \ref{fig:fig-2}B.ii) --- the same total water loss from the same column at the same initial concentration, yet one recedes smoothly and the other intermittently. What differs between these two cases is the drying rate, set by the ambient relative humidity $H_{\mathrm{r}}$. Because efflorescence is nucleation-limited, slower drying allows crystals to appear at lower supersaturation, which lowers $\phi_{\mathrm{eff}}$ and narrows the interval $(\phi_{\mathrm{eff}} - \phi_{\mathrm{sat}})$ between the thresholds (figure \ref{fig:fig-s2}B). A lower threshold at $H_{\mathrm{r}}$ = 0.70 permits the first condition to be met at $\phi_{0}$ = 0.05, where it fails at $H_{\mathrm{r}}$ = 0.10, and the narrower interval gives a smaller concentration excursion per cycle and hence the smaller steps observed there. This is a property of the nucleation-kinetic description (figure \ref{fig:fig-s2}), not of the fixed-threshold model: with $\phi_{\mathrm{eff}}$ held at the value calibrated for $H_{\mathrm{r}}$ = 0.10, the exit concentration does not reach it at $H_{\mathrm{r}}$ = 0.70 within the accessible time window, which is why the model prediction at that humidity shows no precipitation events (figure \ref{fig:fig-s1}C.ii).  \\

It is worth noting what distinguishes this system from evaporation-driven salt growth in other geometries. Growth laws for porous salt structures are commonly derived by treating the dissolved-salt field as quasi-steady \cite{prat-2026-langmuir}. That requires transport in the domain supplying the growth to relax quickly compared to the growth itself. For salt structures growing on a porous substrate that domain is the salt structure itself, and the separation of timescales is roughly two orders of magnitude \cite{prat-2026-langmuir}. Here the domain that supplies the growth is the liquid column inside the capillary, and there is no such timescale separation: $\mathcal{T}_{\mathrm{D}} \approx$ 21 \SI{}{\text{\minute}} against a measured and modeled cycle period of $\approx$ 23 \SI{}{\text{\minute}} at $\phi_{0}$ = 0.20, $H_{\mathrm{r}}$ = 0.10. The concentration field therefore retains memory from one cycle to the next, which is a necessary condition for the unsteady dynamics reported here.  \\

Taken together, this minimal framework demonstrates that intermittent salt creeping and evaporation dynamics emerge from the coupling between one-dimensional advection-diffusion in the liquid column and a hysteretic precipitation-dissolution switch at the salt deposit. The model predicts that oscillations of the salt concentration at the capillary exit between the efflorescence threshold and saturation are both necessary and sufficient to generate intermittent meniscus dynamics, while the failure of either the crossing or the reset yields smooth recession or continuous self-amplified growth. Self-amplified creeping, reported previously for this phenomenon \cite{miri-2015-intjgreenhgascontrol, qazi-2019-sciadv, wijnhorst-2025-arxiv}, is thereby recovered as the limit in which the reset, though still occurring, no longer clears the salt deposit, and the transitions between these dynamical regimes are quantitatively predicted by the theoretical model. By revealing how evaporation-driven transport couples to precipitation and dissolution under confinement, our results recast salt creeping as a genuinely dynamical instability, providing new insight into how phase change and mass transport conspire to generate intermittency far from equilibrium. 

\section{Conclusions and outlook}

We have investigated the confined, unidirectional evaporation of aqueous sodium chloride (\textsf{NaCl}) solutions from one end of a cylindrical capillary and showed that salt creeping on the outer surface of the capillary can fundamentally modify evaporation dynamics. By varying the initial salt mass fraction $\phi_{0}$ and the ambient relative humidity $H_{\mathrm{r}}$, we identified transitions in the evaporation dynamics: from smooth, monotonic meniscus recession to intermittent, step-like dynamics and, close to saturation, to strongly nonlinear acceleration indicating self-amplification. Time-resolved optical measurements revealed that intermittent meniscus jumps are temporally correlated with episodic imbibition of liquid into the porous crystalline network of the creeping salt deposit. This evaporation-imbibition feedback provides an additional pathway for water extraction from the capillary and directly links the dynamics of salt creeping to the observed intermittency. Importantly, salt creeping on its own does not guarantee intermittent dynamics, as creeping may also proceed continuously while the meniscus motion remains smooth. \\

A minimal theoretical model coupling one-dimensional advection-diffusion dynamics to a hysteretic precipitation-dissolution switch at the salt deposit quantitatively captures the observed transitions. The model identifies a simple dynamical criterion: oscillations of the salt concentration at the capillary exit between the efflorescence threshold and saturation generate intermittent activation of salt precipitation and liquid wicking, whereas suppression of these oscillations yields smooth meniscus recession. In this framework, the metastable zone of the salt --- the interval between the supersaturation concentration at which crystals nucleate and the saturation concentration below which they redissolve --- effectively introduces an intrinsic timescale that governs the emergence and disappearance of intermittency under steady external conditions. These results demonstrate that salt creeping is not merely a deposition morphology, but a dynamical process where intermittency emerges when the creeping salt network activates an additional feedback pathway for mass loss from the capillary. In dynamical-systems terms, the confined salt creeper is a relaxation oscillator whose slow \emph{loading} is diffusive transport in the liquid column and whose fast \emph{discharge} is the threshold-triggered crystallization and wicking event. This places salt creeping within the same broad class as the dripping faucet and the saline density oscillator, and suggests that the tools of non-smooth dynamics may predict the boundaries of intermittency in this and related systems. \\

Looking ahead, extending this approach to salts whose metastable zones differ in width \cite{martin-2000-chemrev} will test the generality of the mechanism identified here. Creeping itself is not specific to sodium chloride: it was documented for a range of salts in the earliest reports of the phenomenon \cite{washburn-1926-jphyschem, erlenmeyer-1927-helvchemacta, hazlehurst-1936-jphyschem, huang-1976-nature}, so repeating these measurements for other salts is the natural next step. Efflorescence is fundamentally a nucleation-limited, kinetic transition \cite{desarnaud-2014-jpcl}; replacing the fixed threshold $\phi_{\mathrm{eff}}$ with a supersaturation-dependent nucleation rate would render it an emergent, drying-rate-dependent quantity (figure \ref{fig:fig-s2}), which accounts for the nucleation-controlled suppression of intermittency documented in figure \ref{fig:fig-3}D. What that description does not yet supply is an independent measurement of the barrier to heterogeneous nucleation on the growing deposit, which remains calibrated here. Obtaining one, whether from direct observation of nucleation events or from measurements on deposits of controlled microstructure, would remove the remaining calibrated quantity from the nucleation description. The relaxation oscillator structure identified here --- slow advective-diffusive loading, threshold-triggered discharge, hysteretic reset --- should operate in any confined system where transport couples to a hysteretic phase change, from humidity-cycled nanopores \cite{bellezza-2025-arxiv} to salt-laden building materials under diurnal humidity variation \cite{price-1997-archaeolint, book-goudie, charola-2000-jaminstconserv, linnow-2007-jcultheritage, pel-2004-constbuildmater}. It requires only a nucleation barrier separating the concentration at which crystals form from the concentration at which they dissolve, together with a wicking pathway through the growing deposit. More generally, wherever slow transport loads a system toward a threshold that resets only after the driving has reversed, intermittency of this kind should follow, offering a route to predict and control it in evaporating multicomponent fluids. 

\section{Materials and methods}

\subsection{Experimental protocol}

We have studied the evaporation dynamics of aqueous sodium chloride solutions confined within a cylindrical glass capillary. Solutions were prepared by dissolving sodium chloride (\textsf{NaCl}; ReagentPlus, Sigma-Aldrich) in purified water (Milli-Q). The capillaries (inner diameter 1.0 \SI{}{\text{\milli\metre}}, outer diameter 1.2 \SI{}{\text{\milli\metre}}, length 100 \SI{}{\text{\milli\metre}}; Round Boro Tubing, CM Scientific) were initially filled to a liquid-column height of 30 $\pm$ 2.5 \SI{}{\text{\milli\metre}}. The initial salt mass fraction $\phi_{0}$ was varied between 0.05 and 0.24, spanning a broad range of compositions while remaining below the saturation limit of \textsf{NaCl} in water ($\phi_{\mathrm{sat}}$ = 0.267). \\

To impose unidirectional evaporation, the liquid-filled capillary was mounted vertically between two optically transparent, independently controlled humidity chambers maintained at room temperature. The upper open end of the capillary was exposed to the top chamber, which was held at a relative humidity of 0.90 $\pm$ 0.03, while the lower open end was exposed to the bottom chamber, where the relative humidity $H_{\mathrm{r}}$ was varied between 0.10 $\pm$ 0.03 and 0.70 $\pm$ 0.03. Temperature and humidity were continuously monitored using a calibrated sensor (TSP01, Thorlabs). This configuration ensured that solvent loss occurred exclusively from the lower end of the capillary, thereby enforcing a unidirectional evaporation flux \cite{raju-2024-jfm}. \\

Throughout the experiments the evaporating interface remains spatially fixed at the open lower end of the capillary: the exit stays wetted and evaporation occurs from a liquid surface at the capillary mouth, replenished from the liquid column as water is lost, rather than the interface receding into the capillary bore. It is held there by the sharp circular edge where the open end meets the outer wall of the capillary, at which a contact line is stationary over a wide range of contact angle \cite{oliver-1977-jcis}; the same configuration has been validated for unidirectional evaporation from these capillaries in our previous work \cite{raju-2024-jfm}. As water was lost from the lower end --- through evaporation and, when present, imbibition into creeping salt deposits --- the total length of the liquid column decreased, resulting in a downward displacement of the top meniscus. This displacement, denoted by $y$ (see figure \ref{fig:fig-1}A), provides a direct measurement of the cumulative mass loss from the system \cite{raju-2024-jfm}. \\

Time-resolved measurements were performed by acquiring high-resolution images of the capillary using a digital mirrorless camera (EOS R6 Mark II, Canon) equipped with a macro objective (RF 35 \SI{}{\text{\milli\metre}} F1.8 IS Macro STM, Canon). Images were recorded every 40 \SI{}{\text{\second}} with uniform back-illumination provided by an LED light pad (L4S, Huion). Each experimental condition was repeated at least three times to ensure reproducibility. \\

Post-acquisition image analysis was performed using custom scripts written in Python, employing standard image-processing routines from the OpenCV \cite{opencv} library. Raw greyscale images were converted to binary masks using intensity thresholding, from which the temporal position of the liquid meniscus was extracted. Morphological erosion and dilation operations were subsequently applied to isolate the salt deposits forming on the outer surface of the capillary and to quantify their temporal evolution. \\

Post-mortem scanning electron microscopy of the crystalline deposits on the outer surface of the capillary were performed on a FEI Magellan 400 Scanning Electron Microscope. 

\subsection{Governing equations and boundary conditions}

We introduce a characteristic lengthscale by balancing diffusive transport of salt within the liquid column against diffusive transport of water vapor from the capillary exit to the ambient \cite{book-crank, raju-2024-jfm, martinezpuig-2026-jfm}, $\mathcal{L}_{\mathrm{D}} = \rho \pi \mathcal{D} R / 4 \mathcal{D}_{\mathrm{v}} \, c_{\mathrm{sat}}$, which in turn defines a characteristic timescale, $\mathcal{T}_{\mathrm{D}} = \mathcal{L}_{\mathrm{D}}^{2} / \mathcal{D}$, and a characteristic velocity scale $\mathcal{U}_{\mathrm{D}} = \mathcal{L}_{\mathrm{D}} / \mathcal{T}_{\mathrm{D}}$. Using these scales, we non-dimensionalize the governing equations and boundary conditions, where tildes denote the corresponding non-dimensional variables. \\

The one-dimensional unsteady advection-diffusion equation (equation \eqref{eq:adv-diff}) then reduces to
\begin{align}
    \frac{\partial \phi}{\partial \tilde{t}} + \tilde{u} \, \frac{\partial \phi}{\partial \tilde{z}} = \frac{\partial^{2} \phi}{\partial \tilde{z}^{2}},
    \label{eq:adv-diff-nd}
\end{align}
subject to the initial condition
\begin{align}
    \phi (\tilde{z}, 0) = \phi_{0},
    \label{eq:ic-nd}
\end{align}
while the boundary condition far from the evaporating interface reads
\begin{align}
    \frac{\partial \phi}{\partial \tilde{z}} \bigg \vert_{\tilde{z} \rightarrow \infty} = 0,
    \label{eq:bc-inf-nd}
\end{align}
reflecting the non-evaporating condition at the upper meniscus (figure \ref{fig:fig-4}A). At the evaporating end ($\tilde{z}$ = 0), the conservation of salt mass (equations \eqref{eq:salt-conserv} and \eqref{eq:j-wick}) yields
\begin{align}
    - \tilde{u} \, \phi \vert_{\tilde{z} = 0} + \frac{\partial \phi}{\partial \tilde{z}} = \frac{\tilde{A}_{\mathrm{wick}}}{\tilde{\delta}_{\mathrm{v}}} \left( \chi - H_{\mathrm{r}} \right) \left( \frac{\phi \vert_{\tilde{z} = 0}}{1 - \phi \vert_{\tilde{z} = 0}} \right),
    \label{eq:salt-conserv-nd}
\end{align}
while combining the salt and water mass flux boundary conditions (equations \eqref{eq:water-conserv} -- \eqref{eq:j-wick}) gives
\begin{align}
    \tilde{u} = - \frac{1}{\tilde{\delta}_{\mathrm{v}}} \left( \chi - H_{\mathrm{r}} \right) \left( \frac{\tilde{A}_{\mathrm{wick}}}{1 - \phi \vert_{\tilde{z} = 0}} + \tilde{\delta}_{\mathrm{v}} \right).
    \label{eq:bc-velocity-nd}
\end{align}

The evolution of the wicking surface area is governed by first-order growth and dissolution kinetics (equation \eqref{eq:hysteresis}), which in non-dimensional form become
\begin{align}
    \frac{\mathrm{d} \tilde{A}_{\mathrm{wick}}}{\mathrm{d} \tilde{t}} = \frac{\tilde{A}_{\mathrm{wick}}}{\tilde{\delta}_{\mathrm{v}} \, \tilde{\delta}_{\mathrm{s}}} \left( \chi - H_{\mathrm{r}} \right) \left( \frac{\phi \vert_{\tilde{z} = 0}}{1 - \phi \vert_{\tilde{z} = 0}} \right) \left( \phi \vert_{\tilde{z} = 0} - \phi_{\mathrm{s}} \right),
    \label{eq:hysteresis-nd}
\end{align}
with $(\delta_{\mathrm{s}}, \phi_{\mathrm{s}}) = (\delta_{\mathrm{eff}}, \phi_{\mathrm{sat}})$ in the growth state and $(\delta_{\mathrm{s}}, \phi_{\mathrm{s}}) = (\delta_{\mathrm{sat}}, \phi_{\mathrm{eff}})$ in the dissolution state, the state switching when $\phi \vert_{\tilde{z} = 0}$ last reached $\phi_{\mathrm{eff}}$ or last fell below $\phi_{\mathrm{sat}}$, respectively (see equation \eqref{eq:hysteresis}). \\

The resulting set of coupled, nonlinear equations constitutes an initial-value problem and is solved numerically using an in-house developed Python-based implementation \cite{web-zenodo} based on the LSODA algorithm \cite{hindmarsh-1983, petzold-1983}, which automatically switches between Adams and backward differential formula (BDF) methods to handle stiffness (see Supplementary Information for further details). 

\section{Acknowledgements}

We acknowledge financial support from grant number PID2023-146809OB-I00 funded by the Spanish MICIU/AEI/10.13039/5011000011033 and by ERDF/UE (J.R.R), the Netherlands Organization for Scientific Research (NWO) through the ``Fundamental fluid dynamics challenges in inkjet printing'' (grant number i43; D.L., U.S.), ``Fundamental fluid dynamics challenges in inkjet printing (phase II)'' (grant number KICH2.V4C.20.001; D.L., U.S.), ``ComplexJets'' (grant number KICH2.V4CS.24.001; D.L., U.S.), and ``SuperAssembly'' (grant number OCENW.M.24.177; U.S.), partially co-financed by Canon Production Printing, and the European Research Council (ERC) through the ``DDD'' Advanced Grant (grant number 740479; L.T.R., D.L.). The authors are grateful for the technical assistance from Remco Fokkink and Raoul Fix during the fabrication of the experimental setup, and the Wageningen Electron Microscopy Centre for assistance with scanning electron micsoscopy. The authors thank Alfons van Blaaderen, Christian Diddens, Burak Eral, Valeria Garbin, Steffen Hardt, Thomas Kodger, Marjolein van der Linden, Vatsal Sanjay, Jacco Snoeijer, and Emmanuel Villermaux for insightful discussions. We thank the anonymous referees for their constructive comments and suggestions, which have greatly improved the quality of this manuscript. 

\section{Author contributions}

J.R.R., L.T.R., D.L., and U.S. designed the experimental system. M.M. and L.T.R. performed the experiments and the experimental data analysis. J.R.R. and U.S. developed the theoretical model and performed the associated data analysis. All authors contributed to interpreting the results and writing the manuscript. 

\section{Competing interests}

The authors declare no competing interests.

\section{Data availability}

Data is available from the corresponding author upon request. 


\newpage

\setcounter{equation}{0}
\setcounter{figure}{0}
\setcounter{table}{0}

\renewcommand{\thefigure}{S\arabic{figure}}
\renewcommand{\thetable}{S\arabic{table}}
\renewcommand{\theequation}{S\arabic{equation}}

\section{Supplementary information}

\FloatBarrier

\subsection{Theoretical model quantitatively reproduces the evaporation dynamics}

\begin{figure}
    \centering
    \includegraphics[width=\textwidth]{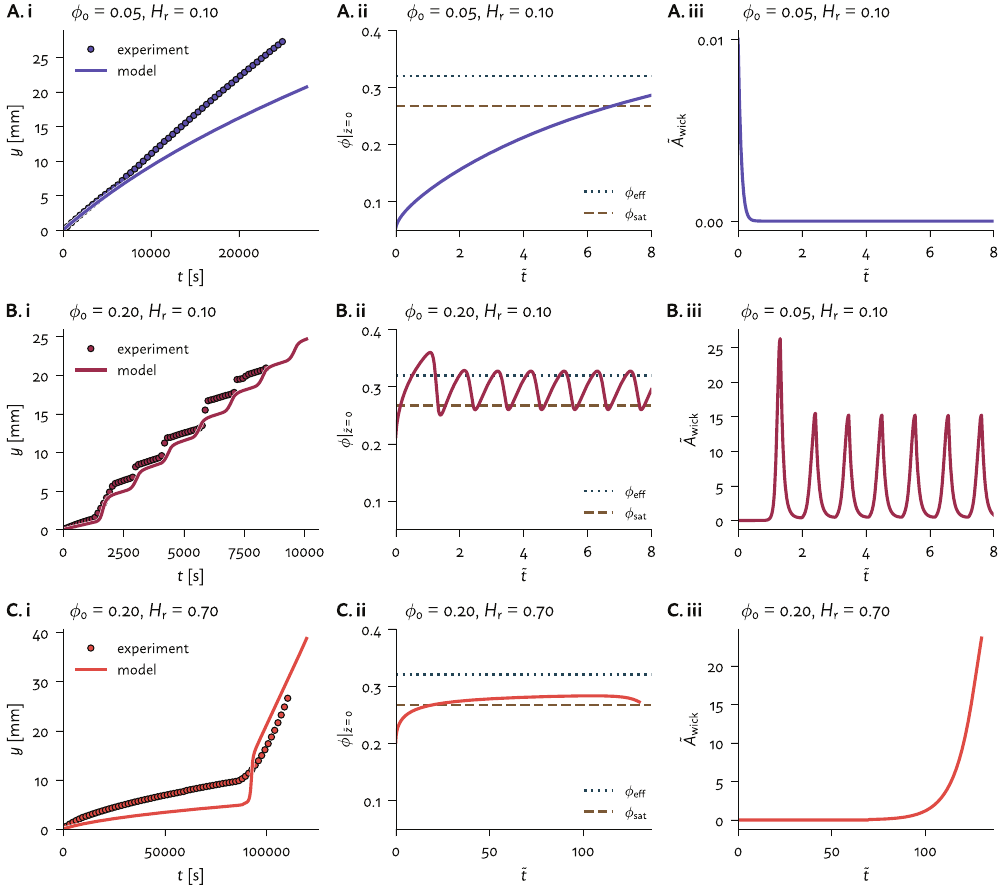}
    \caption{\textbf{Quantitative agreement between experimental evaporation dynamics and theoretical predictions.} Each row is one condition: \textbf{A.} $\phi_{0}$ = 0.05, $H_{\mathrm{r}}$ = 0.10; \textbf{B.} $\phi_{0}$ = 0.20, $H_{\mathrm{r}}$ = 0.10; \textbf{C.} $\phi_{0}$ = 0.20, $H_{\mathrm{r}}$ = 0.70. For each row: \textbf{i.} Top-meniscus displacement $y (t)$ in dimensional units, theoretical prediction (solid line) against experimental measurements (discrete markers); only a selection of experimental datapoints is shown to avoid overcrowding of the plot. \textbf{ii.} Exit concentration $\phi \vert_{\tilde{z} = 0}$, with $\phi_{\mathrm{sat}}$ (dashed line) and $\phi_{\mathrm{eff}}$ (dotted line), and \textbf{iii.} wicking area $\tilde{A}_{\mathrm{wick}}$, both against non-dimensional time $\tilde{t}$.  }
    \label{fig:fig-s1}
\end{figure}

A quantitative comparison between the experimentally measured evaporation dynamics and the predictions of the theoretical model for the non-salt-creeping regime ($H_{\mathrm{r}}$ = 0.10 and $\phi_{0}$ = 0.05; figure \ref{fig:fig-1}B.ii and movie SM1) is shown in figure \ref{fig:fig-s1}A.i. Prior to the onset of crystallization ($t \approx$ 6680 s, temporally correlated with an upward kink in the experimentally-measured recession dynamics shown in figure \ref{fig:fig-s1}A.i), the model shows excellent agreement with the experimental measurements of the meniscus recession. Following nucleation, the model underpredicts the experimentally observed recession dynamics. This discrepancy arises from the minimal nature of the theoretical framework, which does not explicitly account for crystal nucleation or for changes in the microstructure of the deposited salt that subsequently imbibes liquid. \\

The agreement in this \emph{smooth} regime is obtained with no fitted transport coefficients. The vapor-transport scale is therefore fixed by the material properties alone, independently of the hysteresis parameters that govern the intermittent regime. We note that alternative choices for the microscopic length scales $\delta_{\mathrm{v}}$, $\delta_{\mathrm{eff}}$, and $\delta_{\mathrm{sat}}$ could improve quantitative agreement between theory and experiment. However, these parameters are associated with molecular-scale transport and interfacial processes that are not directly measurable in the present experiments. We therefore refrain from further parameter fitting and focus instead on the mechanistic insights provided by the model. \\

With the calibration described later in the Supplementary Information, the minimal theoretical model additionally reproduces the intermittent regime ($\phi_{0}$ = 0.20, $H_{\mathrm{r}}$ = 0.10; figures \ref{fig:fig-1}B.i and figure \ref{fig:fig-s1}B.i, and movie SM2). Started in the loading phase so that the first discharge coincides with the physical onset in the experiments, the model shows reasonable agreement with the experiments for the timing of the individual step-like jumps of $y (t)$, the time period between the jumps, and the amplitude of each individual jump (figure \ref{fig:fig-s1}B.i). The temporal variations of the exit concentration $\phi \vert_{\tilde{z} = 0}$ (figure \ref{fig:fig-s1}B.ii) and the wicking area in $\tilde{A}_{\mathrm{wick}}$ (figure \ref{fig:fig-s1}B.iii) make the relay directly visible: the exit concentration climbs to $\phi_{\mathrm{eff}}$, the wicking area jumps, the exit is flushed and falls, and the area relaxes as the finest asperities dissolve. The step structure of $y (t)$ is the integrated consequence of that switching rather than an independent feature. The mismatch between the model predictions and the experiment in figure \ref{fig:fig-s1}B.i reflects the genuine cycle-to-cycle irregularity of the experiment, which the fixed-parameter model does not reproduce. \\

At elevated humidity ($\phi_{0}$ = 0.20, $H_{\mathrm{r}}$ = 0.70; figures \ref{fig:fig-2}A.i and figure \ref{fig:fig-s1}C.i, and movie SM4), with $\phi_{\mathrm{eff}}$ held at the value calibrated for $H_{\mathrm{r}}$ = 0.10, the exit concentration settles into a quasi-steady balance between evaporative removal and diffusive resupply that lies below the threshold: it reaches only $\approx$ 0.29 and no precipitation event occurs (figure \ref{fig:fig-s1}C.ii). The experiment nucleates at $t \approx$ 87640 \SI{}{\text{\second}} regardless, because the threshold is itself lower at high humidity, as the nucleation-kinetic description in figure \ref{fig:fig-s2}B makes quantitative. To compare model and experiment in this regime we therefore seed the crust at the measured nucleation time ($t \approx$ 87640 \SI{}{\text{\second}}, figure \ref{fig:fig-s1}C.i) and let the relay take over. The onset is an input taken from the measurement; the subsequent dynamics --- the acceleration of the recession and the growth of the wicking area --- is what the model predicts, and it reproduces the observed self-amplification.    

\subsection{Nucleation-limited efflorescence}

\begin{figure}
    \centering
    \includegraphics[width=\textwidth]{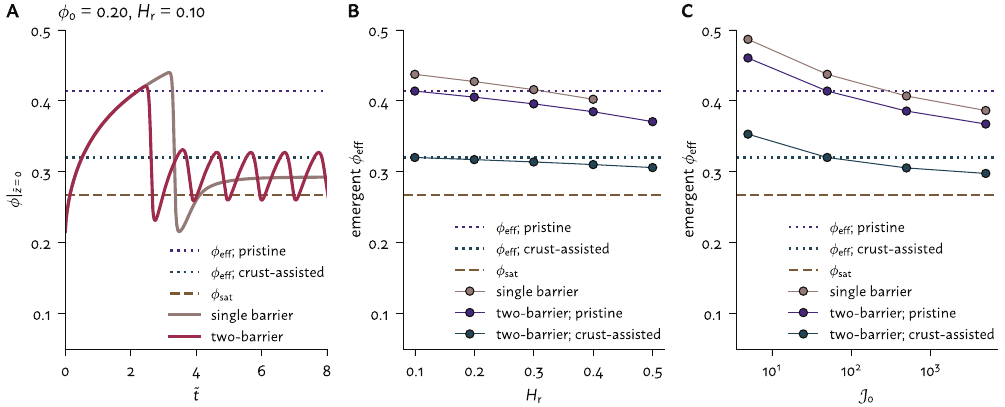}
    \caption{\textbf{Efflorescence as a nucleation-limited transition.} The fixed efflorescence threshold $\phi_{\mathrm{eff}}$ of the model presented in the main text (equation \eqref{eq:hysteresis-nd}) is replaced by a classical-nucleation theory (CNT) rate, which shows that the intermittent mechanism is robust and that $\phi_{\mathrm{eff}}$ is an emergent, nucleation-controlled quantity. \textbf{A.} Exit concentration $\phi \vert_{\tilde{z} = 0}$ against time for the single-barrier variant and the two-barrier closure, at $\phi_{0}$ = 0.20 and $H_{\mathrm{r}}$ = 0.10. The single-barrier variant nucleates once and then stalls, because the wicking area is held fixed through the metastable zone; the two-barrier closure, in which the finest asperities continue to dissolve between events, sustains cycling. Emergent $\phi_{\mathrm{eff}}$ against \textbf{B.} the ambient humidity $H_{\mathrm{r}}$ and \textbf{C.} the nucleation prefactor $\mathcal{J}_{0}$ for the single-barrier variant and for both branches of the two-barrier closure: \emph{pristine}, which governs the first precipitation event, and \emph{crust-assisted}, which governs the repeated cycle. Dotted lines: $\phi_{\mathrm{eff}}$, dashed line: $\phi_{\mathrm{sat}}$. The crust-assisted branch reproduces the calibrated value at $H_{\mathrm{r}}$ = 0.10 without having been fitted to it, while the single-barrier curve tracks the pristine branch --- which is why it nucleates once rather than cycling. The interval $( \phi_{\mathrm{eff}} - \phi_{\mathrm{sat}} )$, which sets the size of each precipitation event, is the gap from each curve down to the $\phi_{\mathrm{sat}}$ line. }
    \label{fig:fig-s2}
\end{figure}

In the minimal theoretical model presented in the main manuscript, the efflorescence threshold $\phi_{\mathrm{eff}}$ is a fixed parameter: crust re-forms whenever the exit concentration reaches $\phi_{\mathrm{eff}}$. Physically, however, efflorescence is not a fixed thermodynamic threshold but a kinetic, nucleation-limited transition --- crystals appear at a supersaturation that depends on the drying rate, the available nucleation sites, and time \cite{tang-1994-jgeophysres, martin-2000-chemrev, mikhailov-2009-atmoschemphys, desarnaud-2014-jpcl}. Dissolution, by contrast, is a genuine thermodynamic transition occurring at a well-defined water activity, which is why the saturation threshold $\phi_{\mathrm{sat}}$ is treated as a material constant while $\phi_{\mathrm{eff}}$ must be calibrated. To test that the intermittent-creeping mechanism does not depend on treating efflorescence as a fixed threshold, and to connect $\phi_{\mathrm{eff}}$ to measurable nucleation physics, we replace the constant $\phi_{\mathrm{eff}}$ by a classical-nucleation-theory (CNT) estimation. \\

In this variant, efflorescence is a stochastic nucleation process with a nucleation rate given by \cite{martin-2000-chemrev, book-kaschiev, book-kelton}
\begin{align}
    \mathcal{J} (\mathcal{S}) = \mathcal{J}_{0} \exp \left( - \frac{\mathcal{B}}{\ln^{2} \mathcal{S}} \right),
\end{align}
where $\mathcal{J}_{0}$ is a rate prefactor, $\mathcal{B}$ a dimensionless nucleation-barrier coefficient, and the supersaturation ratio at the capillary exit, $\mathcal{S}$, given by
\begin{align}
    \mathcal{S} = \frac{\phi \vert_{\tilde{z} = 0} / \left( 1 - \phi \vert_{\tilde{z} = 0} \right)}{\phi_{\mathrm{sat}} / \left( 1 - \phi_{\mathrm{sat}} \right)}.
    \label{eq:si-supersat}
\end{align}
The rate vanishes below saturation and rises steeply as supersaturation builds; the efflorescence event is triggered when the expected number of nuclei, $n (\tilde{t}) = \int \mathcal{J} \, \mathrm{d} \tilde{t}^{\prime}$, reaches unity --- the mean-field limit of a Poisson nucleation process --- after which the counter is reset \cite{book-debenedetti, binder-1987-rpp, goh-2010-crystgrowthdes}. The remaining dynamics is unchanged from the baseline model: the growth branch is identical, and the quiescent branch holds the crust in the metastable zone while the nuclei accumulate. Equation \eqref{eq:si-supersat} also fixes the meaning of the metastability limit quoted in the literature. The limiting supersaturation \cite{desarnaud-2014-jpcl} $\mathcal{S}^{\ast}$ is the same quantity as $\mathcal{S}$, evaluated where crystals first appear in a clean, deposit-free capillary under slow drying; for \textsf{NaCl} in confinement \cite{desarnaud-2014-jpcl} $\mathcal{S}^{\ast} \approx$ 1.6. With $\phi_{\mathrm{sat}}$ = 0.267 this corresponds to a mass fraction of 0.368, whereas the calibrated threshold of the main text, $\phi_{\mathrm{eff}}$ = 0.320, corresponds to $\mathcal{S} \approx$ 1.29. The calibrated threshold therefore lies below the metastability limit, as it must, since crystallization in the sustained cycle occurs on an existing deposit rather than in clean solution.  \\

A single barrier coefficient, however, does not sustain multi-cycle intermittency (figure \ref{fig:fig-s2}A), and the reason is instructive. In the single-barrier implementation the wicking area is held fixed throughout the metastable zone, on the grounds that the deposit is inert while the solution is supersaturated. Frozen at the large value left by the first event, it continues to draw solution from the exit, whose concentration stalls at 0.292 --- a supersaturation of only $\mathcal{S} \approx$ 1.13 --- and never re-approaches $\phi_{\mathrm{eff}}$. At that supersaturation a second event would require a time interval of order 350 $\mathcal{T}_{\mathrm{D}}$, against an experimental time window of 8--20 $\mathcal{T}_{\mathrm{D}}$. Lowering the barrier does not help, because the exit never reaches the supersaturation at which the barrier height matters. The resolution is physical rather than numerical: the finest crystalline asperities, which are the features that carry the wicking, have a solubility elevated above the bulk value by the Gibbs-Thomson effect \cite{book-mullin}. An asperity of radius $a$ is in equilibrium with a solution whose supersaturation ratio, defined as in equation \eqref{eq:si-supersat}, is
\begin{align}
    \mathcal{S}_{a} = \exp \left( \mathscr{L}_{\mathrm{GT}} / a \right) \, ,
\end{align}
with
\begin{align}
    \mathscr{L}_{\mathrm{GT}} = \frac{2 \gamma \Omega}{\nu k_{\mathrm{B}} T} \, ,
\end{align}
where $\gamma$ is the crystal-solution interfacial free energy, $\Omega$ the molar volume of the solid, $\nu$ = 2 the number of ions per formula unit, $k_{\mathrm{B}}$ the Boltzmann constant, and $T$ the absolute temperature. For \textsf{NaCl} at room temperature, $\gamma \approx$ 0.11 \SI{}{\text{\joule.\meter\tothe{-2}}} and $\Omega \approx$ 4.5 $\times$ 10$^{\text{- 29}}$ \SI{}{\text{\meter\tothe{3}}}, giving a Gibbs-Thomson capillary length $\mathscr{L}_{\mathrm{GT}} \approx$ 1.2 \SI{}{\text{\nano\meter}}. An asperity therefore continues to dissolve whenever the exit supersaturation satisfies $\mathcal{S} < \mathcal{S}_{a}$, i.e. whenever $a < \mathscr{L}_{\mathrm{GT}} / \ln \mathcal{S}$. Immediately above saturation this cutoff is large --- $\approx$ 79 \SI{}{\text{\nano\meter}} at $\mathcal{S}$ = 1.015 --- and it falls steeply as the exit loads: $\approx$ 18 \SI{}{\text{\nano\meter}} at $\mathcal{S}$ = 1.07, and $\approx$ 5 \SI{}{\text{\nano\meter}} by the time the exit reaches the threshold at $\mathcal{S}$ = 1.29. The finest features of the deposit therefore keep dissolving and coarsening through the early part of the metastable zone, even though the solution is already supersaturated with respect to the bulk salt, and the relaxation shuts off progressively as loading proceeds. This is what allows the wicking area to relax between events \cite{scherer-1999-cemconcrres, steiger-2005-jcrystgrowth, steiger-2005-jcrysgrowth-2}. Restoring that relaxation --- the same dissolution law the relay of the main text already uses --- allows the kinetic variant to cycle indefinitely (figure \ref{fig:fig-s2}A).  \\

With the wicking area free to relax, the description also distinguishes two barriers that physically should not be equal. \emph{Pristine} nucleation, in a clean capillary, is a rare, high-barrier event; its coefficient $\mathcal{B}_{\mathrm{p}}$ is fixed by requiring that a slow ramp nucleate at the metastability limit $\mathcal{S}^{\ast} \approx$ 1.6, giving $\mathcal{B}_{\mathrm{p}}$ = 1.37. \emph{Crust-assisted} nucleation, on the deposit left by the previous cycle, proceeds at smaller supersaturation since the existing crystal surfaces on the capillary walls lower the barrier \cite{prat-2002-chemengj, shahidzadeh-2008-langmuir}; its coefficient $\mathcal{B}_{\mathrm{c}} < \mathcal{B}_{\mathrm{p}}$ governs every subsequent cycle and is the single calibrated parameter of this description, replacing the $\phi_{\mathrm{eff}}$ of the main text. The parameter count is therefore unchanged: one calibrated barrier in place of one calibrated threshold, with $\mathcal{J}_{0}$ = 50 throughout. A caveat to note: the supersaturation at which the model actually nucleates is set by the drying rate and not by $\mathcal{S}^{\ast}$ itself: under the drying conditions of the experiments the exit passes $\mathcal{S}^{\ast}$ before enough nuclei have accumulated, and the first, pristine event occurs at $\mathcal{S}$ = 1.94. This is not an inconsistency but the central point of what follows --- a faster-driven system nucleates at higher supersaturation than a quasi-static one.  \\

With the two barriers so assigned, the loading-burst-reset cycle survives, but the efflorescence threshold is now an output rather than an input: the exit concentration at which nucleation emerges from the competition between rising supersaturation and the accumulating nucleation probability (figure \ref{fig:fig-s2}A). Because nucleation is kinetic, the emergent threshold depends on the drying rate --- a slower drying rate allows nucleation to occur closer to its intrinsic limit --- so $\phi_{\mathrm{eff}}$ becomes systematically lower as the ambient humidity increases (figure \ref{fig:fig-s2}B). This drying-rate dependence, absent from the fixed-threshold model, is a direct signature of the kinetic nature of efflorescence and connects the confined-capillary system to the aerosol efflorescence literature, where the same phenomenology is observed \cite{tang-1994-jgeophysres, mikhailov-2009-atmoschemphys}. The same calculation explains why the fixed-threshold model does not nucleate at $H_{\mathrm{r}}$ = 0.70: the crust-assisted threshold has fallen to 0.306 by $H_{\mathrm{r}}$ = 0.50 and continues to fall, toward the $\approx$ 0.29 that the exit concentration actually reaches, which is why the comparison in figure \ref{fig:fig-s1}C.i requires a seeded crust. What matters dynamically, however, is not the threshold itself but the interval $(\phi_{\mathrm{eff}} - \phi_{\mathrm{sat}})$, which sets the concentration excursion available to each cycle. Across the humidity range of figure \ref{fig:fig-s2}B the threshold falls by $\approx$ 4\% while the interval falls by $\approx$ 27\%, because the interval is a small difference between two comparable numbers and amplifies relative changes in the threshold by a factor $\phi_{\mathrm{eff}} / (\phi_{\mathrm{eff}} - \phi_{\mathrm{sat}}) \approx$ 6. A modest kinetic effect therefore has a large dynamical consequence, and this is why slower drying produces visibly smaller steps, as observed at $\phi_{0}$ = 0.05, $H_{\mathrm{r}}$ = 0.70 (figure \ref{fig:fig-2}B.ii).  \\

The same variant explains the nucleation-control experiment of the main text (figure \ref{fig:fig-3}D and movie SM6) mechanistically. Increasing the nucleation prefactor $\mathcal{J}_{0}$ --- physically, providing more heterogeneous nucleation sites --- lowers the emergent threshold toward saturation (figure \ref{fig:fig-s2}C), and with it the interval that powers the cycle (figure \ref{fig:fig-s2}C); raising $\mathcal{J}_{0}$ by three decades lowers the threshold from 0.354 to 0.298 and narrows the interval from 0.087 to 0.031, a reduction of about two thirds. As nucleation occurs before significant supersaturation can build, the interval $(\phi_{\mathrm{eff}} - \phi_{\mathrm{sat}})$ collapses and the oscillation is suppressed. This is precisely the observed effect of roughening the capillary base: the added nucleation sites correspond to a larger $\mathcal{J}_{0}$, and the intermittency disappears. Efflorescence being nucleation-controlled, any manipulation that eases nucleation lowers the effective threshold and narrows the interval that powers the oscillation. Plotted together, the single-barrier curve tracks the pristine branch across both sweeps, which is why it nucleates once rather than sustaining a cycle: it describes nucleation in a clean capillary throughout, whereas the repeated cycle occurs on an existing deposit. The two-barrier closure also recovers the $\phi_{\mathrm{eff}}$ used in the main text. Calibrating $\mathcal{B}_{\mathrm{c}}$ gives an emergent crust-assisted threshold at $H_{\mathrm{r}}$ = 0.10 of 0.320, against 0.320 obtained independently in the main text from the measured oscillation period and step timings. The fixed threshold of the main text is therefore not an independent assumption but the effective heterogeneous threshold that nucleation kinetics, anchored to an independently measured metastability limit, produce for the repeated cycle.  \\

For these reasons, we retain the fixed-threshold model for all quantitative results in the main text and use the kinetic variant only to establish the mechanism and its robustness. The kinetic description requires two barrier coefficients and a rate prefactor, none of which can be measured independently in the present experiments, in place of one calibrated number that is pinned by a direct observable --- the oscillation period and step timing. Adopting it throughout would replace a well-constrained fit with an under-constrained one and weaken exactly the quantitative claims that are strongest. Its value is therefore explanatory rather than quantitative: it establishes that the intermittent mechanism survives making efflorescence kinetic, that $\phi_{\mathrm{eff}}$ is fundamentally an emergent, drying-rate-dependent quantity, and that the nucleation-control experiment follows from the same mechanism --- thereby fixing the physical interpretation of the effective threshold used in the main text. 

\subsection{P\'{e}clet number as a model output}

\begin{figure}
    \centering
    \includegraphics[width=\textwidth]{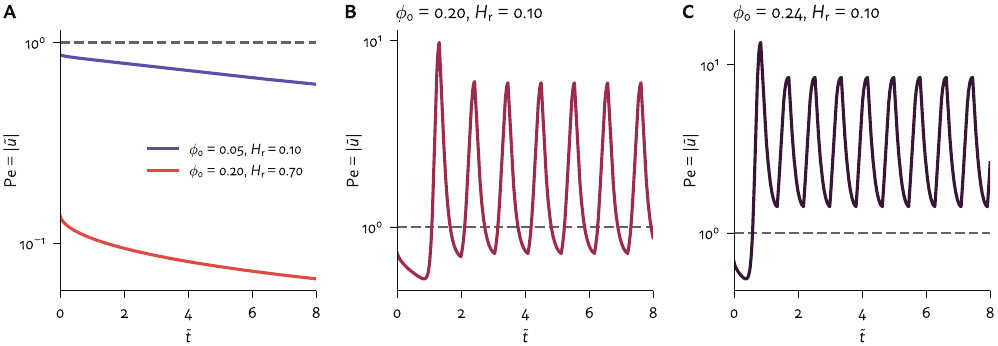}
    \caption{\textbf{The P\'{e}clet number at the capillary exit is an output of the model.} $\mathrm{Pe}$ = $\vert \tilde{u} \vert$ against non-dimensional time $\tilde{t}$; dashed horizontal line indicates $\mathrm{Pe}$ = 1. \textbf{A.} Conditions in which the exit concentration does not reach $\phi_{\mathrm{eff}}$ within the time window: $\mathrm{Pe}$ decreases monotonically and is diffusive ($\mathrm{Pe} < 1$) throughout. \textbf{B.} The intermittent condition, $\phi_{0}$ = 0.20 at $H_{\mathrm{r}}$ = 0.10: each event raises $\tilde{A}_{\mathrm{wick}}$ abruptly (figure \ref{fig:fig-s1}B.iii) and drives $\mathrm{Pe}$ an order of magnitude above its loading value, so the exit alternates once per cycle between diffusive loading ($\mathrm{Pe} <$ 1) and advective discharge ($\mathrm{Pe} >$ 1). \textbf{C.} The self-amplified condition, $\phi_{0}$ = 0.24 at $H_{\mathrm{r}}$ = 0.10: the cycle persists, but the exit barely dips below $\phi_{\mathrm{sat}}$ (figure \ref{fig:fig-s4}A), $\tilde{A}_{\mathrm{wick}}$ ratchets upward from cycle to cycle (figure \ref{fig:fig-s4}B), and after the first cycle the loading intervals no longer return below $\mathrm{Pe}$ = 1. } 
    \label{fig:fig-s3}
\end{figure}

Since lengths and times are non-dimensionalized with $\mathcal{L}_{\mathrm{D}}$ and $\mathcal{T}_{\mathrm{D}} = \mathcal{L}_{D}^{2} / \mathcal{D}$, the velocity scale is $\mathcal{U}_{\mathrm{D}} = \mathcal{D} / \mathcal{L}_{\mathrm{D}}$, and the P\'{e}clet number of the exit region, $\mathrm{Pe} = \vert u \vert \, \mathcal{L}_{\mathrm{D}} / \mathcal{D}$, is identically $\vert \tilde{u} \vert$, the magnitude of the non-dimensional advection velocity of equation \eqref{eq:bc-velocity-nd}. It is therefore an output of the model rather than a quantity that has to be assumed, and is available at every instant for every condition, with no additional parameter. The variation of $\mathrm{Pe} (\tilde{t})$, grouped by dynamical regime, is shown in figure \ref{fig:fig-s3}.  \\

Where the exit concentration does not reach $\phi_{\mathrm{eff}}$, $\mathrm{Pe}$ decreases monotonically and never approaches unity (figure \ref{fig:fig-s3}A). With $\phi_{\mathrm{eff}}$ held at the value calibrated for $H_{\mathrm{r}}$ = 0.10, neither of the two conditions ($\phi_{0}$ = 0.05, $H_{\mathrm{r}}$ = 0.10 and $\phi_{0}$ = 0.20, $H_{\mathrm{r}}$ = 0.70) reached the threshold within the explored time window, which is the same limitation discussed in relation to figure \ref{fig:fig-s1}; both are therefore shown here as conditions in which transport remains diffusion-controlled ($\mathrm{Pe} < 1$) throughout.    \\

The intermittent condition ($\phi_{0}$ = 0.20, $H_{\mathrm{r}}$ = 0.10) alternates between the two transport regimes (advective: $\mathrm{Pe} > 1$ and diffusive: $\mathrm{Pe} < 1$) once per cycle (figure \ref{fig:fig-s3}B). Each crystallization event raises $\tilde{A}_{\mathrm{wick}}$ abruptly (figure \ref{fig:fig-s4}B) and drives $\mathrm{Pe}$ to 9.8, and order of magnitude above its loading value; between events $\mathrm{Pe}$ relaxes to 0.70, so the loading phase is diffusion-controlled and the discharge advection-controlled. This is the transport-level statement of the same relay that governs the meniscus: slow diffusive loading, fast advective discharge.  \\

The self-amplified condition ($\phi_{0}$ = 0.24, $H_{\mathrm{r}}$ = 0.10) is distinguished by where the cycle sits relative to $\mathrm{Pe}$ = 1 (figure \ref{fig:fig-s3}C). At $\phi_{0}$ = 0.24 the exit concentration still crosses both thresholds repeatedly, but it barely dips below $\phi_{\mathrm{sat}}$ (figure \ref{fig:fig-s4}A), so the unsaturated interval is too short for $\tilde{A}_{\mathrm{wick}}$ to relax (figure \ref{fig:fig-s4}B), and the wicking area ratchets upward from cycle to cycle, ending the run some eightfold larger than at $\phi_{0}$ = 0.20. After the first cycle the loading intervals no longer return to the diffusive regime: $\mathrm{Pe}$ rests at 1.44 between events and peaks at 13.5 (figure \ref{fig:fig-s3}C). Transport at the exit is then advection-dominated at all times, which is the transport signature of continuous, self-amplified growth, and the reason the meniscus recession appears smooth despite an underlying cycle that has not stopped. The minimum of $\mathrm{Pe}$ during the settled loading intervals therefore separates the two regimes without reference to the shape of $y (t)$: 0.70 for the intermittent condition, 1.44 for the self-amplified one. 

\subsection{The reset distinguishes intermittent from self-amplified creeping}

\begin{figure}
    \centering
    \includegraphics[width=\textwidth]{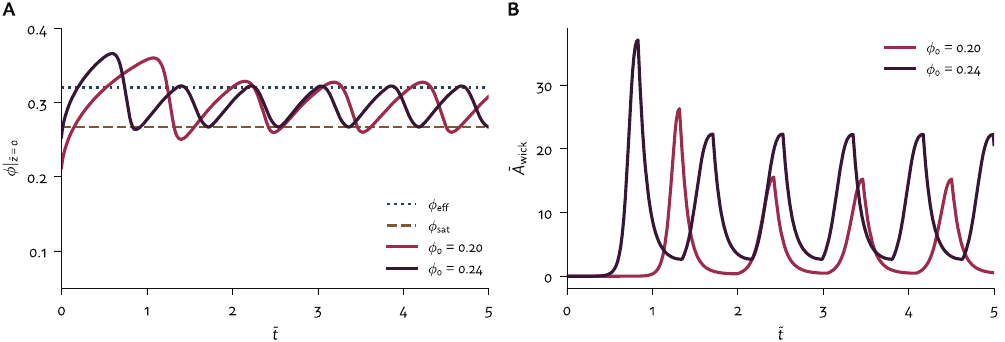}
    \caption{\textbf{The reset distinguishes intermittent from self-amplified creeping.} \textbf{A.} Exit concentration (dashed line: $\phi_{\mathrm{sat}}$; dotted line: $\phi_{\mathrm{eff}}$) and \textbf{B.} wicking area at $\phi_{0}$ = 0.20 (intermittent) and $\phi_{0}$ = 0.24 (self-amplified), both at $H_{\mathrm{r}}$ = 0.10, against non-dimensional time $\tilde{t}$. Both conditions cycle between the thresholds. At $\phi_{0}$ = 0.20 each discharge carries the exit below $\phi_{\mathrm{sat}}$, while at $\phi_{0}$ = 0.24 the exit dips barely below $\phi_{\mathrm{sat}}$, and the unsaturated interval is too short for $\tilde{A}_{\mathrm{wick}}$ to relax.   }
    \label{fig:fig-s4}
\end{figure}

Both the intermittent and the self-amplified conditions cycle. What separates them is not whether the exit concentration crosses the thresholds but how completely the reset undoes the growth, and figure S4 shows this directly for $\phi_{0}$ = 0.20 and 0.24 at $H_{\mathrm{r}}$ = 0.10.  \\

At $\phi_{0}$ = 0.20 each discharge carries the exit well below saturation --- to 0.250 against $\phi_{\mathrm{sat}}$ = 0.267 --- and the unsaturated interval is long enough for the finest asperities to dissolve (figure \ref{fig:fig-s4}A). The wicking area returns to a floor of about 0.44 before the next event (figure \ref{fig:fig-s4}B), so each cycle begins close to where the last one began and the salt deposit grows in discrete, comparable steps.  \\

At $\phi_{0}$ = 0.24 the bulk reservoir is itself within 0.027 of saturation, so the solution resupplied to the exit is almost saturated. The exit still marginally falls below $\phi_{\mathrm{sat}}$, but only to 0.264 (figure \ref{fig:fig-s4}A), and the unsaturated interval is too short for the wicking area to relax: its floor settles near 2.67, six times higher, and it ends the run at 20.7 against 0.55 at $\phi_{0}$ = 0.20 (figure \ref{fig:fig-s4}B). The deposit therefore ratchets upward from cycle to cycle, and the recession becomes effectively continuous. Therefore, the self-amplified regime is not the absence of the cycle but the failure of the reset to clear the deposit. 

\subsection{Numerical procedure and model calibration}

The governing equations for the minimal theoretical model are integrated numerically using a finite difference method. The liquid column is discretized on a uniform grid of $N$ points over a truncated domain $0 \leq \tilde{z} \leq \tilde{z}_{\infty}$ with a zero-gradient (Neumann) condition at the far boundary. Spatial derivatives use second-order central differences in the interior and a three-point one-sided stencil at the exit. At each evaluation, the exit concentration $\phi \vert_{\tilde{z} = 0}$, is obtained by solving the implicit salt-flux boundary condition (a scalar root-find). \\

The coupled system for the wick area, meniscus position, and concentration field is advanced with an adaptive, stiffness-switching LSODA integrator \cite{web-zenodo, hindmarsh-1983, petzold-1983}. Since the crust dynamics is a two-state relay, each branch (growth or dissolution) is integrated until a terminal event --- the exit concentration crossing the relevant threshold ($\phi_{\mathrm{eff}}$ or $\phi_{\mathrm{sat}}$) --- at which point the state switches and integration resumes from the event. This event-driven segmentation resolves the discontinuous switching exactly rather than smoothing it. Two safeguards ensure robustness near the switching manifold: the wicking area is floored at a small positive value (it is an invariant manifold of the dynamics and cannot become negative), and event localization is retried at a reduced maximum step if the bracketing root-find fails on a grazing threshold crossing. Simulations are performed using $N$ = 400 grid points, $\tilde{z}_{\infty}$ = 40, and maximum time step of order 0.02$\mathcal{T}_{\mathrm{D}}$. The model predictions are insensitive to grid refinement at these resolutions. \\

The minimal theoretical model has a small number of physical parameters, almost all of which are set either by known values in literature (for $\mathcal{D}$, $\mathcal{D}_{\mathrm{v}}$, $\phi_{\mathrm{sat}}$, $c_{\mathrm{sat}}$, and $\chi (H_{\mathrm{r}})$) or by experimental design ($R$, $H_{\mathrm{r}}$, and $\phi_{0}$). The smooth regime recession rate, measured before the appearance of the first crystal (figure \ref{fig:fig-s1}A.i), is captured almost identically by the model by only setting these values, and requires no further calibration or free parameters. The intermittent and self-amplifying recession dynamics require the precipitation-dissolution switch parameters ($\delta_{\mathrm{v}}$ = 1 \SI{}{\text{\milli\meter}}, $\delta_{\mathrm{eff}}$ = $\delta_{\mathrm{sat}}$ = 0.9 \SI{}{\text{\micro\meter}}, and $\phi_{\mathrm{eff}}$ = 0.320), which are calibrated against three independent experimental measurements for the $\phi_{0}$ = 0.20, $H_{\mathrm{r}}$ = 0.10 case. With these values, the model predictions agree reasonably well with the experimental measurements (figures \ref{fig:fig-s1}B.i and \ref{fig:fig-s1}C.i). 

\subsection{Alternative mechanisms}

Four mechanisms other than the precipitation-dissolution cycle could in principle produce intermittency here. We set out briefly why each is inconsistent with our measurements.  \\

Solutal Marangoni flow \cite{marin-2019-prf, lohse-2020-natrevphys} acts at the meniscus inside the capillary and is continuous rather than thresholded. It offers no mechanism for the observed temporal correlation between the meniscus steps and optical changes in the deposit on the \emph{outer} wall (figure \ref{fig:fig-3}A.i), nor for the suppression of intermittency when nucleation at the capillary base is enhanced (figure \ref{fig:fig-3}D), since roughening the base does not alter the internal interface.   \\

A sustained osmotic pressure difference requires a barrier selectively permeable to solvent over solute; the deposit is an open porous medium through which both are transported, and salt is carried through it and re-precipitated at its outer surface. The ionic effects \cite{book-israelachvili} that could produce an effective selectivity --- double-layer overlap, electroviscous retardation, Donnan exclusion --- all require the Debye length to be comparable to the pore size. For a near-saturated \textsf{NaCl} solution the Debye length is $\approx$ 0.12 \SI{}{\text{\nano\meter}} against a pore space of order 1--100 \SI{}{\text{\micro\meter}} (figure \ref{fig:fig-3}B), a factor of 10$^{\text{4}}$--10$^{\text{6}}$, so none is available.  \\

Liesegang-type periodic precipitation \cite{nabika-2020-langmuir} requires two counter-diffusing reactants and a reaction between them; the present system has a single solute and no reaction. It also predicts spatially periodic banding of the deposit, which a cycle at a moving front need not produce. We have further examined the optical images and the scanning electron micrographs of figure \ref{fig:fig-3}B and find no such banding.  \\

Stick-slip depinning \cite{watanabe-2025-pre} of the evaporating interface would itself produce step-like $y (t)$. A depinning event that lets the interface retreat into the capillary bore would occur once and irreversibly, and would register as a step change in the evaporation rate, not as repeated bounded steps. The steps are locked to greyscale value dips in the deposit on the outer wall (figure \ref{fig:fig-3}A.i), which such an event cannot produce; and roughening the base (figure \ref{fig:fig-3}D), which alters the wetting conditions substantially, makes the recession smoother rather than more stick-slip. 

\subsection{Supplementary movies}

\textbf{Movie SM1:} Confined evaporation of aqueous sodium chloride solution at $\phi_{0}$ = 0.05 and $H_{\mathrm{r}}$ = 0.10.  \\

\textbf{Movie SM2:} Confined evaporation of aqueous sodium chloride solution at $\phi_{0}$ = 0.20 and $H_{\mathrm{r}}$ = 0.10. \\

\textbf{Movie SM3:} Confined evaporation of aqueous sodium chloride solution at $\phi_{0}$ = 0.24 and $H_{\mathrm{r}}$ = 0.10. \\

\textbf{Movie SM4:} Confined evaporation of aqueous sodium chloride solution at $\phi_{0}$ = 0.20 and $H_{\mathrm{r}}$ = 0.70.  \\

\textbf{Movie SM5:} Confined evaporation of aqueous sodium chloride solution at $\phi_{0}$ = 0.05 and $H_{\mathrm{r}}$ = 0.70.  \\

\textbf{Movie SM6:} Confined evaporation of aqueous sodium chloride solution at $\phi_{0}$ = 0.20 and $H_{\mathrm{r}}$ = 0.10 with roughened base. 


\FloatBarrier

\newpage

\bibliographystyle{unsrt}
\bibliography{capillary+salt}


\end{document}